\documentclass[a4paper,11pt]{article}
\pdfoutput=1 

\usepackage{jheppub} 
\usepackage[T1]{fontenc}
\usepackage{gensymb}
\def\be{\begin{equation}}
\def\ee{\end{equation}}
\newcommand{\com}{} 

\usepackage{lipsum}

\newcommand\blfootnote[1]{%
  \begingroup
  \renewcommand\thefootnote{}\footnote{#1}%
  \addtocounter{footnote}{-1}%
  \endgroup
}

\title{\boldmath State-dependent regulation of cortical processing speed via gain modulation}

\author[a]{David Wyrick}
\author[a,b]{and Luca Mazzucato}
\affiliation{${}^a$ Department of Biology and Institute of Neuroscience, ${}^b$Department of Mathematics and Physics, University of Oregon, Eugene.}
\emailAdd{lmazzuca at uoregon dot edu}

\abstract{To thrive in dynamic environments, animals must be capable of rapidly and flexibly adapting behavioral responses to a changing context and internal state. Examples of behavioral flexibility include faster stimulus responses when attentive and slower responses when distracted. Contextual or state-dependent modulations may occur early in the cortical hierarchy and may be implemented via top-down projections from cortico-cortical or neuromodulatory pathways. However, the computational mechanisms mediating the effects of such projections are not known. Here, we introduce a theoretical framework to classify the effects of cell-type specific top-down perturbations on the information processing speed of cortical circuits. Our theory demonstrates that perturbation effects on stimulus processing can be predicted by intrinsic gain modulation, which controls the timescale of the circuit dynamics. Our theory leads to counter-intuitive effects such as improved performance with increased input variance. We tested the model predictions using large-scale electrophysiological recordings from the visual hierarchy in freely running mice, where we found that a decrease in single-cell intrinsic gain during locomotion led to an acceleration of visual processing. Our results establish a novel theory of cell-type specific perturbations, applicable to top-down modulation as well as optogenetic and pharmacological manipulations. Our theory links connectivity, dynamics, and information processing via gain modulation.
}

\begin{document} 
\maketitle
\flushbottom

\section{Introduction}
\label{sec:intro}

Animals respond to the same stimulus with different reaction times depending on the context or the behavioral state. Faster responses may be elicited by expected stimuli or when the animal is aroused and attentive \cite{niemi1981foreperiod}. Slower responses may occur in the presence of distractors or when the animal is disengaged from the task \cite{grueninger1969effects,treisman1980feature,desimone1995neural}. Experimental evidence suggests that neural correlates of these contextual modulations occur early in the cortical hierarchy, already at the level of the primary sensory cortex \cite{jaramillo2010auditory,samuelsen2012effects}. During the waking state, levels of arousal, attention, and task engagement vary continuously and are associated with ongoing and large changes in the activity of neuromodulatory systems \cite{lee2014identification,pinto2013fast,fu2014cortical} as well as cortico-cortical feedback pathways \cite{guo2014flow,chen2017map,nelson2013circuit,leinweber2017sensorimotor,zhang2014long}. Activation of these pathways modulate the patterns of activity generated by cortical circuits and may affect their information-processing capabilities. However, the precise computational mechanism underlying these flexible reorganizations of cortical dynamics remains elusive.

Variations in behavioral and brain state, such as arousal, engagement and body movements may act on a variety of timescales, both slow (minutes, hours) and rapid (seconds or subsecond), and spatial scales, both global (pupil diameter, orofacial movements) and brain subregion-specific; and they can be recapitulated by artificial perturbations \com{such as optogenetic, chemogenetic or electrical stimulation}. These variations have been associated with a large variety of seemingly unrelated mechanisms operating both at the single cell and at the population level. At the population level, these mechanisms include modulations of low and high frequency rhythmic cortical activities \cite{mcginley2015waking}; changes in noise correlations \cite{cohen2009attention,dadarlat2017locomotion}; and increased information flow between cortical and subcortical networks \cite{mcginley2015waking}. On a cellular level, these variations have been associated with modulations of single-cell responsiveness and reliability \cite{dadarlat2017locomotion};  and cell-type specific gain modulation \cite{mcginley2015waking}. These rapid, trial-by-trial modulations of neural activity may be mediated by neuromodulatory pathways, such as cholinergic and noradrenergic systems \cite{lee2014identification,pinto2013fast,fu2014cortical,reimer2016pupil}, or more precise cortico-cortical projections from prefrontal areas towards primary sensory areas \cite{guo2014flow,chen2017map,nelson2013circuit,leinweber2017sensorimotor,zhang2014long}. The effects of these cortico-cortical projections can be recapitulated by optogenetic activation of glutamatergic feedback pathways \cite{zagha2015competing}. In the face of this wide variety of physiological pathways, is there a common computational principle underlying the effects they elicit on sensory cortical circuits? 

A natural way to model the effect of activating a specific pathway on a downstream circuit is in the form of a perturbation to the downstream circuit's afferent inputs or recurrent couplings \cite{mazzucato2019expectation,huang2019circuit}. 
Here, we will present a theory explaining how these perturbations control the information-processing speed of a downstream cortical circuit. Our theory shows that the effects of perturbations that change the statistics of the afferents or the recurrent couplings can all be captured by a single mechanism of action: intrinsic gain modulation, \com{where gain is defined as the rate of change of the intrinsic input/output transfer function of a neuron measured during periods of ongoing activity.} Our theory is based on a biologically plausible model of cortical circuits using clustered spiking network \cite{Amit1997b}. This class of models capture complex physiological properties of cortical dynamics such as state-dependent changes in neural activity, variability \cite{,LitwinKumarDoiron2012,DecoHugues2012,mazzucato2015dynamics,mazzucato2016stimuli,rostami2020spiking} and information-processing speed \cite{mazzucato2019expectation}. Our theory predicts that gain modulation controls the \com{intrinsic} temporal dynamics of the cortical circuit and thus its information processing speed, such that decreasing the intrinsic single-cell gain leads to faster stimulus coding. 

We tested our theory by examining the effect of locomotion on visual processing in the visual hierarchy. We found that locomotion decreased the intrinsic gain of \com{visual cortical} neurons in the absence of stimuli in freely running mice. The theory thus predicted a faster encoding of visual stimuli during running compared to rest, which we confirmed in the empirical data. Our theoretical framework links gain modulation to information-processing speed, providing guidance for the design and interpretation of future manipulation experiments by unifying the changes in brain state due to behavior, optogenetic, or pharmacological perturbations, under the same shared mechanism.

\section{Methods}
\subsection{Spiking network model} 
{\it Architecture}. We modeled the local cortical circuit as a network of $N=2000$ excitatory (E) and inhibitory (I) neurons (with relative fraction $n_E=80\%$ and $n_I=20\%$) with random recurrent connectivity (Fig.~\ref{figtwo}). Connection probabilities were $p_{EE}=0.2$ and $p_{EI}=p_{IE}=p_{II}=0.5$. Nonzero synaptic weights from pre-synaptic neuron $j$ to post-synaptic neuron $i$ were $J_{ij}={j_{ij}/\sqrt{N}}$, with $j_{ij}$ sampled from a gaussian distribution with mean $j_{\alpha\beta}$, for $\alpha,\beta=E,I$, and standard deviation $\delta^2$. E and I neurons were arranged in $p$ clusters. E clusters had heterogeneous sizes drawn from a gaussian distribution with a mean of $N^{clust}_E=80$ E-neurons and $20\%$ standard deviation. The number of clusters was then determined as $p=\textrm{round}(n_EN(1-n_{bgr})/N^{clust}_E)$, where $n_{bgr}=0.1$ is the fraction of background neurons in each population, i.e., not belonging to any cluster. I clusters had equal size $N^{clust}_I=\textrm{round}( n_IN(1-n_{bgr}/p)$. \com{Clusters were defined by an increase in intra-cluster weights and a decrease in inter-cluster weights, under the constraint that the net input current to a neuron would remain unchanged compared to the case without clusters.} Synaptic weights for within-cluster neurons where potentiated by a ratio factor $J^{+}_{\alpha\beta}$. Synaptic weights between neurons belonging to different clusters were depressed by a factor $J^{-}_{\alpha\beta}$. Specifically, we chose the following scaling: $J^{+}_{EI}=p/(1+(p-1)/g_{EI})$, $J^{+}_{IE}=p/(1+(p-1)/g_{IE})$, $J^-_{EI}=J^{+}_{EI}/g_{EI}$, $J^-_{IE}=J^{+}_{IE}/g_{IE}$ and
$J^-_{\alpha\alpha}=1-\gamma (J^+_{\alpha\alpha}-1)$ for $\alpha=E,I$, with $\gamma=f(2-f(p+1))^{-1}$, where $f=(1-n_{bgr})/p$ is the fraction of E neurons in each cluster. Within-cluster E-to-E synaptic weights were further multiplied by cluster-specific factor equal to the ratio between the average cluster size $N^{clust}_E$ and the size of each cluster, so that larger clusters had smaller within-cluster couplings. We chose network parameters so that the cluster timescale was 100 ms, as observed in cortical circuits \cite{Jones2007,mazzucato2015dynamics,mazzucato2019expectation}. Parameter values are in Table \ref{table:1}.

\begin{table}[!t]
\begin{tabular}{ |p{2cm}|p{10cm}||p{2cm}|  }
\hline
\multicolumn{3}{|c|}{Model parameters for clustered network simulations} \\
\hline
Parameter& Description & Value \\
\hline
$j_{EE}$ & mean E-to-E synaptic weights $\times\sqrt{N}$ & 0.6 mV\\
$j_{IE}$ & mean E-to-I synaptic weights $\times\sqrt{N}$&0.6 mV\\
$j_{EI}$ &  mean I-to-E synaptic weights $\times\sqrt{N}$&1.9 mV\\ 
$j_{II}$ & mean I-to-I synaptic weights $\times\sqrt{N}$ &3.8 mV\\
$j_{E0}$ & mean E-to-E synaptic weights $\times\sqrt{N}$ &2.6 mV\\
$j_{I0}$ & mean I-to-I synaptic weights $\times\sqrt{N}$ &2.3 mV\\
$\delta$ & standard deviation of the synaptic weight distribution&$20\%$\\
$J^{+}_{EE}$ & Potentiated intra-cluster E-to-E weight factor &14\\
$J^{+}_{II}$ & Potentiated intra-cluster I-to-I weight factor &5\\
$g_{EI}$ & Potentiation parameter for intra-cluster I-to-E weights &10\\
$g_{IE}$ & Potentiation parameter for intra-cluster E-to-I weights &8\\
$r_{ext}$ & Average baseline afferent rate to E and I neurons &5 spks/s\\
$V^{thr}_E$ & E-neuron threshold potential &1.43 mV\\
$V^{thr}_I$ & I-neuron threshold potential &0.74 mV\\
$V^{reset}$ & E- and I-neuron reset potential &0 mV\\
$\tau_m$ & E- and I-neuron membrane time constant  &20 ms\\
$\tau_{refr}$ & E- and I-neuron absolute refractory period  &5 ms\\
$\tau_s$ & E- and I-neuron synaptic time constant  &5 ms\\
\hline
\end{tabular}
\caption{Parameters for the clustered network used in the simulations.}
\label{table:1}
\end{table}

\noindent{\it Neuronal dynamics}. We modeled spiking neurons as current-based leaky-integrate-and-fire (LIF) neurons whose membrane potential $V$ evolved according to the dynamical equation
$$
{dV\over dt}=-{V\over \tau_m} +I_{rec} +I_{ext} \ ,
$$
where $\tau_m$ is the membrane time constant. Input currents included  a contribution $I_{rec}$ coming from the other recurrently connected neurons in the local circuit and an external
current $I_{ext} = I_0 + I_{stim} + I_{pert}$ (units of mV s${}^{-1}$). The first term $I_0=N_{ext}J_{\alpha0}r_{ext}$ (for $\alpha=E,I$) is a constant term representing input to the E or I neuron from other brain areas and $N_{ext}=n_ENp_{EE}$; while $I_{stim}$ and $I_{pert}$ represent the incoming sensory stimulus or the various types of perturbation (see Stimuli and perturbations below). When $V$ hits threshold $V^{thr}_{\alpha}$ (for $\alpha=E,I$), a spike
is emitted and $V$ is then held at the reset value $V^{reset}$ for a refractory period $\tau_{refr}$. We chose the thresholds so that the homogeneous network (i.e.,where all $J^{\pm}_{\alpha\beta}=1$) was in a balanced state with average spiking activity at rates
$(r_E,r_I) = (2,5)$ spks/s \cite{Amit1997b,mazzucato2019expectation}. \com{Post-synaptic currents} evolved according to the following equation
$$
\tau_{syn}{dI_{rec}\over dt}=-I_{rec}+\sum_{j=1}^NJ_{ij}\sum_{k}\delta(t-t_k) \ ,
$$
where $\tau_s$ is the synaptic time constant, $J_{ij}$ are the recurrent couplings and $t_k$ is the time of the k-th spike from the
j-th presynaptic neuron. Parameter values are in Table \ref{table:1}. 

\begin{figure*}[!t]
\vspace{-0.5cm}
\includegraphics[width=1\linewidth]{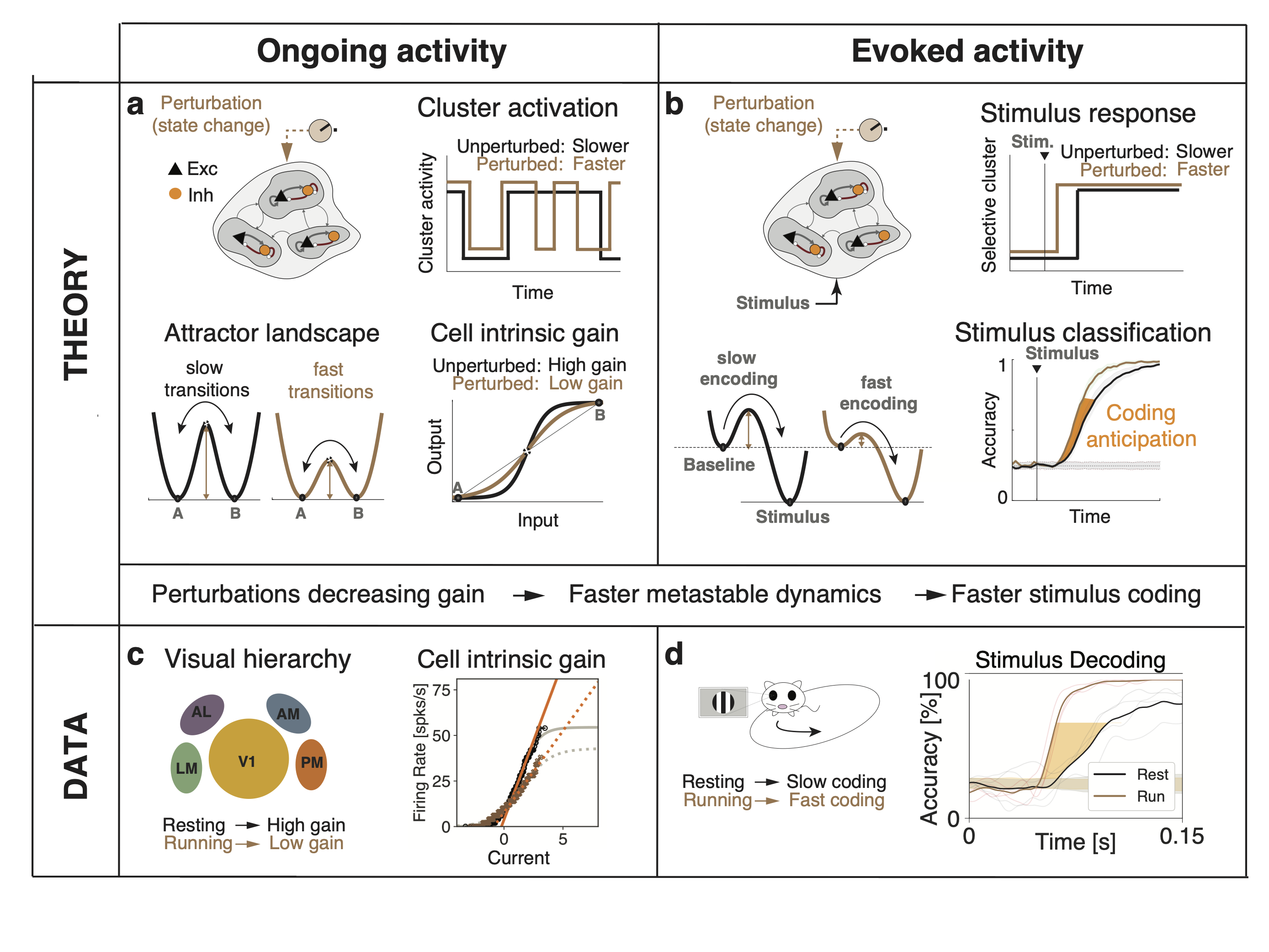}
\vspace{-0.5cm}
\caption{Conceptual summary of the main results. {\bf a)}: In a network model of sensory cortex featuring clusters of excitatory and inhibitory neurons with metastable dynamics, state changes are induced by external perturbations controlling the timescale of cluster activation during ongoing activity. The neural mechanism underlying timescale modulation is a change in the barrier height separating attractors, driven by a modulation of the intrinsic gain of the single-cell transfer function. {\bf b)}: During evoked activity, onset of stimulus encoding is determined by the activation latency of stimulus-selective cluster. External perturbations modulate the onset latency thus controlling the stimulus processing speed. The theory shows that the effect of external perturbations on stimulus-processing speed during evoked activity (right) can be predicted by the induced gain modulations observed during ongoing activity (left). {\bf c)}: Locomotion induced changes in intrinsic gain in the visual cortical hierarchy during darkness periods. {\bf d)}: Locomotion drove faster coding of visual stimuli during evoked periods, as predicted by the induced gain modulations observed during ongoing activity.
}
\label{figone}
\end{figure*}

{\it \com{Sensory stimuli}}. We considered two classes of inputs: sensory stimuli and perturbations. In the ``evoked'' condition (\com{Fig. \ref{figfour}a}), we presented the network \com{one of four sensory} stimuli, modeled as changes in the afferent currents targeting $50\%$ of E-neurons in stimulus-selective clusters; each E-cluster had a $50\%$ probability of being selective to a sensory stimulus (mixed selectivity). \com{In the first part of the paper (Fig. \ref{figone}-\ref{figsix}, I-clusters were not stimulus-selective}. Moreover, in both the unperturbed and the perturbed stimulus-evoked conditions, stimulus onset occurred at time $t=0$ and each stimulus was represented by an afferent current $I_{stim}(t)=I_{ext}r_{stim}(t)$, where $r_{stim}(t)$ is a linearly ramping increase reaching a value $r_{max}=20\%$ above baseline at $t = 1$. \com{In the last part of the paper (Fig. \ref{figseven}), we introduced a new stimulation protocol where visual stimuli targeted both E and I clusters in pairs\com{, corresponding to thalamic input onto both excitatory and inhibitory neurons in V1 \cite{zhuang2013layer,miska2018sensory,kloc2014target,khan2018distinct}.}. Each E-I cluster pair had a 50\% probability of being selective to each visual stimulus. If a E-I cluster pair was selective to a stimulus, then all E neurons and 50\% of I neurons in that pair received the stimulus. The time course of visual stimuli was modeled as a double exponential profiles with rise and decay times of (0.05,0.5)s, and peak equal to a 20\% increase compared to the baseline external current.}

{\it \com{External perturbations}}. We considered several kinds of perturbations. In the perturbed stimulus-evoked condition (\com{Fig. \ref{figfour}b}, right panel), perturbation onset occurred at time $t=-0.5$ and lasted until the end of the stimulus presentation at $t=1$ with a constant time course. We also presented perturbations in the absence of sensory stimuli (``ongoing'' condition, \com{Fig. \ref{figtwo}-\ref{figthree}}); in that condition, the perturbation was constant and lasted for the whole duration of the trial (5s). Finally, when assessing single-cell responses to perturbations, we modeled the perturbation time course as a double exponential with rise and decay times $[0.1,1]$s (\com{Fig. \ref{figsix}}). In all conditions, perturbations were defined as follows:
\begin{itemize}
    \item $\delta$mean(E), $\delta$mean(I): A constant offset $I_{pert}=zI_0$ in the mean afferent currents was added to all neurons in either E or I populations, respectively, expressed as a fraction of the baseline value $I_0$ (see Neuronal dynamics above), where $z\in[-0.1,0.2]$ for E neurons and $z\in[-0.2,0.2]$ for I neurons.
    \item $\delta$var(E), $\delta$var(I): For each E or I neuron, respectively, the perturbation was a constant offset $I_{pert}=zI_0$, where $z$ is a gaussian random variable with zero mean and standard deviation $\sigma$. We chose $\sigma\in[0,0.2]$ for E neurons and $\sigma\in[0,0.5]$ for I neurons. This perturbation did not change the mean afferent current but only its spatial variance across the E or I population, respectively. We measured the strength of these perturbations via their coefficient of variability $CV(\alpha)=\sigma_\alpha/\mu_\alpha$, for $\alpha=E,I$, where $\sigma$ and $\mu=I_0$ are the standard deviation and mean of the across-neuron distribution of afferent currents.
    \item $\delta$AMPA: A constant change in the mean $j_{\alpha E}\to(1+z)j_{\alpha E}$ synaptic couplings (for $\alpha=E,I$), representing a modulation of glutamatergic synapses. We chose $z\in[-0.1,0.2]$.    
    \item $\delta$GABA: A constant change in the mean $j_{\alpha I}\to(1+z)j_{\alpha I}$ synaptic couplings (for $\alpha=E,I$), representing a modulation of GABAergic synapses. We chose $z\in[-0.2,0.2]$.
\end{itemize}
The range of the perturbations were chosen so that the network still produced metastable dynamics for all values.

{\it Inhibition stabilization}. We simulated a stimulation protocol used in experiments to test inhibition stabilization (\com{Fig. \ref{figsone}b}). This protocol is identical to the $\delta$mean(I) perturbation during ongoing periods, where the perturbation targeted all I neurons with an external current $I_{pert}=zI_0$ applied for the whole length of 5s intervals, with $z\in[0,1.2]$ and  40 trials per network and 10 networks for each value of the perturbation. 

{\it Simulations}. All data analyses, model simulations, and mean field theory calculations were performed using custom software written in MATLAB, C and Python. Simulations in the stimulus-evoked conditions (both perturbed and unperturbed) comprised 10 realizations of each network (each network with different realization of synaptic weights), with 20 trials for each of the 4 stimuli. Simulations in the ongoing condition comprised 10 different realization of each network, with 40 trials per perturbation. Each network was initialized with random synaptic weights
and simulated with random initial conditions in each trial. Sample
sizes were similar to those reported in previous publications \cite{mazzucato2015dynamics,mazzucato2016stimuli,mazzucato2019expectation}. Dynamical equations for the leaky-integrate-and-fire neurons were integrated with the Euler method with a 0.1ms step. MATLAB code to simulate the model with $\delta$var(E) perturbation is located at \url{https://github.com/mazzulab/perturb_spiking_net}. Code to reproduce the full set of perturbations investigated in this paper are available upon request to the corresponding author.

\begin{table}[!t]
\begin{tabular}{ |p{2cm}|p{10cm}||p{2cm}|  }
\hline
\multicolumn{3}{|c|}{Model parameters for the reduced two-cluster network} \\
\hline
Parameter& Description & Value \\
\hline
$j_{EE}$ & mean E-to-E synaptic weights $\times\sqrt{N}$ & 0.8 mV\\
$j_{EI}$ &  mean I-to-E synaptic weights $\times\sqrt{N}$&10.6 mV\\ 
$j_{IE}$ & mean E-to-I synaptic weights $\times\sqrt{N}$&2.5 mV\\
$j_{II}$ & mean I-to-I synaptic weights $\times\sqrt{N}$ &9.7 mV\\
$j_{E0}$ & mean E-to-E synaptic weights $\times\sqrt{N}$ &14.5 mV\\
$j_{I0}$ & mean I-to-I synaptic weights $\times\sqrt{N}$ &12.9 mV\\
$J^{+}_{EE}$ & Potentiated intra-cluster E-to-E weight factor &11.2\\
$r^{ext}$ & Average baseline afferent rate to E and I neurons &7 spk/s\\
$V^{thr}_E$ & E-neuron threshold potential &4.6 mV\\
$V^{thr}_I$ & I-neuron threshold potential &8.7 mV\\
$\tau_s$ & E- and I-neuron synaptic time constant  &4 ms\\
$n_{bgr}$ & Fraction of background E neurons  &65$\%$\\
\hline
\end{tabular}
\caption{Parameters for the simplified two-cluster network used for the mean-field theory analysis (the remaining parameters are in Table \ref{table:1}.}
\label{table:2}
\end{table}

\subsection{Mean field theory}

We performed a mean field analysis of a simplified two-cluster network for leaky-integrate-and-fire neurons with exponential synapses, comprising $p+2$ populations for $p=2$ \cite{Amit1997b,mazzucato2019expectation}: the first $p$ representing the two E clusters, the last two representing the background E and the I population. The infinitesimal mean $\mu_n$ and variance $\sigma_n^2$ of the postsynaptic currents are:
\begin{eqnarray}
\label{eq:mean}
    \mu_n&=&\tau_m\sqrt{N}\left[n_Ep_{EE}j_{EE}\left(fJ^+_{EE}r_n+J_{EE}^-(\sum_{l=1}^{p-1}r_l+(1-pf)r_E^{bgr})+{j_{E0}\over j_{EE}}r_{ext}\right)-n_{I}p_{EI}j_{EI}r_I\right] \ , \nonumber\\
    \mu_{bgr}&=&\tau_m\sqrt{N}\left[n_Ep_{EE}j_{EE}\left(J_{EE}^-\sum_{l=1}^{p}r_l+(1-pf)r_E^{bgr}+{j_{E0}\over j_{EE}}r_{ext}\right)-n_{I}p_{EI}j_{EI}r_I\right] \ , \nonumber\\    
    \mu_I&=&\tau_m\sqrt{N}\left[n_Ep_{IE}j_{IE}\left(f\sum_{l=1}^{p}r_l+(1-pf)r_E^{bgr}\right)-n_{I}p_{II}(j_{II}r_I+j_{I0}r_{ext})\right] \ , 
\end{eqnarray}
\begin{eqnarray}
\label{eq:variance}
    \sigma^2_n&=&\tau_m\sqrt{N}\left[n_Ep_{EE}j_{EE}^2\left(f(J^+_{EE})^2r_n+(J^-_{EE})^2(\sum_{l=1}^{p-1}r_l+(1-pf)r_E^{bgr}))\right)-n_{I}p_{EI}j_{EI}^2r_I\right] \ , \nonumber\\    
    \sigma^2_{bgr}&=&\tau_m\sqrt{N}\left[n_Ep_{EE}j^2_{EE}\left((J_{EE}^-)^2\sum_{l=1}^{p}r_l+(1-pf)r_E^{bgr}\right)-n_{I}p_{EI}j^2_{EI}r_I\right] \ , \nonumber\\    
    \sigma_I&=&\tau_m\sqrt{N}\left[n_Ep_{IE}j^2_{IE}\left(f\sum_{l=1}^{p}r_l+(1-pf)r_E^{bgr}\right)-n_{I}p_{II}j^2_{II}r_I\right] \ ,     
\end{eqnarray}
where $r_n,r_l=1,\ldots,p$ are the firing rates in the $p$ E-clusters; $r_E^{bgr},r_I,r_{ext}$ are the firing rates in the background E population, in the I population, and in the external current. Other parameters are described in Architecture and in Table \ref{table:2}. The network attractors satisfy the self-consistent fixed point equations:
\begin{equation}
\label{eq:fixedpoints}
    r_{l}=F_l[\mu_l({\bf r}),\sigma_l^2({\bf r})] \ ,
\end{equation}
where ${\bf r}=(r_1,\ldots,r_p,r_{bgr},r_I)$ and $l=1,\ldots,p,bgr,I$, and $F_l$ is the current-to-rate transfer function for each population, which depend on the condition. In the absence of perturbations, all populations have the LIF transfer function
\begin{equation}
\label{eq:transfer}
    F_l(\mu_l,\sigma_l)=\left(\tau_{refr}+\tau_m\sqrt{\pi}\int_{H_l}^{\Theta_l}e^{u^2}[1+\textrm{erf}(u)]\right)^{-1} \ ,
\end{equation}
where $H_l=(V^{reset}-\mu_l)/\sigma_l+ak$ and $\Theta_l=(V^{thr}_l-\mu_l)/\sigma_l+ak$. $k=\sqrt{\tau_s/\tau_m}$ and $a=|\zeta(1/2)|/\sqrt{2}$ are terms accounting for the synaptic dynamics \cite{Fourcaud2002}. The perturbations $\delta$var(E) and $\delta$var(I) induced an effective population transfer function $F^{eff}$ on the E and I populations, respectively, given by \cite{mazzucato2019expectation}:
\begin{equation}
\label{eq:transferEff}
    F^{pert}_\alpha(\mu_\alpha,\sigma_\alpha)=\int{\cal}DzF_\alpha(\mu_\alpha+z\sigma_z\mu^{ext}_\alpha,\sigma^2_\alpha) \ ,
\end{equation}
where $\alpha=E,I$ and ${\cal}Dz=dz\,\exp(-z^2/2/\sqrt{2\pi})$ is a gaussian measure of zero mean and unit variance, $\mu^{ext}_\alpha=\tau_m\sqrt{N}n_\alpha p_{\alpha0}j_{\alpha0}r_{ext}$ is the external current and $\sigma_z$ is the standard deviation of the perturbation with respect to baseline, denoted CV(E) and CV(I). Stability of the fixed point equation \ref{eq:fixedpoints} was defined with respect to the approximate linearized dynamics of the instantaneous mean $m_l$ and variance $s_l^2$ of the input currents \cite{mazzucato2015dynamics,mazzucato2019expectation}:
\begin{equation}
    \tau_s{dm_l\over dt}=-m_l+\mu_l(r_l) \ ; \quad \tau_s{ds^2_l\over 2dt}=-s^2_l+\sigma^2_l(r_l) \ ; \quad r_l=F_l(m_l({\bf r}),s^2_l({\bf r})) \ ,
\end{equation}
where $\mu_l,\sigma^2_l$ are defined in \ref{eq:mean}-\ref{eq:variance} and $F_l$ represents the appropriate transfer function \ref{eq:transfer} or \ref{eq:transferEff}. Fixed point stability required that the stability matrix
\begin{equation}
\label{eq:stability}
    S_{lm}={1\over \tau_s}\left({\partial F_l(\mu_l,\sigma^2_l)\over \partial r_m}-{\partial F_l(\mu_l,\sigma^2_l)\over \partial \sigma^2_l}{\partial\sigma^2_l({\bf r})\over \partial r_m}-\delta_{lm} \right) \ ,
\end{equation}
was negative definite. The full mean field theory described above was used for the comprehensive analysis of Fig. \ref{figsthree}.For the schematic of Fig. \ref{figthree}c, we replaced the LIF transfer function \ref{eq:transfer} with the simpler function $\tilde F(\mu_E) =0.5(1+\tanh(\mu_E))$ and the $\delta$var(E) perturbation effect was then modeled as $\tilde F^{eff}(\mu) =\int{ D}z\tilde F(\mu_E+z\sigma_z\mu_{ext})$.

{\it Effective mean field theory for a reduced network}. To calculate the potential energy barrier separating the two network attractors in the reduced two-cluster network, we used the effective mean field theory developed in \cite{Mascaro1999,mattia2013heterogeneous,mazzucato2019expectation}. The idea is to first estimate the force acting on neural configurations with cluster firing rates ${\bf r}=[\tilde r_1, \tilde r_2]$ outside the fixed points (\ref{eq:fixedpoints}), then project the two-dimensional system onto a one-dimensional trajectory along which the force can be integrated to give an effective potential $E$ (Fig. \ref{figsthree}). In the first step, we start from the full mean field equations for the $P=p+2$ populations in \ref{eq:fixedpoints}, and obtain an effective description of the dynamics for $q$ populations ``in focus'' describing E clusters ($q=2$ in our case) by integrating out the remaining $P-q$ out-of-focus populations describing the background E neurons and the I neurons ($P-q=2$ in our case). Given a fixed value $\tilde {\bf r} = [\tilde r_1 ,\ldots, \tilde r_q ]$ for the $q$ in-focus populations, one obtains the stable fixed point firing rates ${\bf r}'=[r'_{q+1},\ldots,r'_{P}]$ of the out-of-focus populations by solving their mean field equations
\begin{equation}
\label{eq:outoffocus}
    {r'_\beta}(\tilde {\bf r})=F_\beta[\mu_\beta(\tilde{\bf r},{\bf r}'),\sigma^2_\beta(\tilde{\bf r},{\bf r}')] \ ,
\end{equation}
for $\beta=q+1,\ldots,P$, as function of the in-focus populations $\tilde {\bf r}$, where stability is calculated with respect to the condition (\ref{eq:stability}) for the reduced $(q+1,\ldots,P)$ out-of-focus populations at fixed values of the in-focus rates $\tilde {\bf r}$. One then obtains a relation between the input $\tilde {\bf r}$ and output values $\tilde {\bf r}^{out}$ of the in-focus populations by inserting the fixed point rates of the out-of-focus populations calculated in (\ref{eq:outoffocus}):
\begin{equation}
\label{eq:infocus}
    {r^{out}_\alpha}(\tilde {\bf r})=F_\alpha[\mu_\alpha(\tilde{\bf r},{\bf r}'(\tilde{\bf r})),\sigma^2_\alpha(\tilde{\bf r},{\bf r}'(\tilde{\bf r}))] \ ,
\end{equation}
for $\alpha=1,\ldots,q$. The original fixed points are $\tilde {\bf r}^*$ such that $\tilde r_\alpha^*=r^{out}_\alpha(\tilde {\bf r^*})$.

{\it Potential energy barriers and transfer function gain.} In a reduced network with two in-focus populations $[\tilde r_1,\tilde r_2]$ corresponding to the two E clusters, one can visualize Eq. (\ref{eq:infocus}) as a two-dimensional force vector $\tilde {\bf r}-{\bf r}^{out}(\tilde {\bf r})$ at each point in the two-dimensional firing rate space $\tilde {\bf r}$. The force vanishes at the stable fixed points A and B and at the unstable fixed point C between them (Fig. \ref{figsthree}). One can further reduce the system to one dimension by approximating its dynamics along the trajectory between A and B as \cite{Mascaro1999}:
\begin{equation}
\label{onedim}
    {\tau_s}{d\tilde r\over dt}=-\tilde r+r^{out}(\tilde r) \ ,
\end{equation}
where $y=r^{out}(\tilde r)$ represents an effective transfer function and $\tilde r-r^{out}(\tilde r)$ an effective force. We estimated the gain  $g$  of the effective transfer function as $g=1-{r^{out}(\tilde r_{min})-r^{out}(\tilde r_{min})\over \tilde r_{min}-\tilde r_{max}}$, where $\tilde r_{min}$ and $\tilde r_{max}$ represent, respectively, the minimum and maximum of the force (see Fig. \ref{figsthree}). From the one-dimensional dynamics (\ref{onedim}) one can define a potential energy via ${\partial E(\tilde r)\over \partial r}=\tilde r-r^{out}(\tilde r)$. The energy minima represent the stable fixed points $A$ and $B$ and the saddle point $C$ between them represents the potential energy barrier separating the two attractors. The height $\Delta$ of the potential energy barrier is then given by
\begin{equation}
    \Delta=\int_{A}^Cd\tilde r[\tilde r-r^{out}(\tilde r)] \ ,
\end{equation}
which can be visualized as the area of the curve between the effective transfer function and the diagonal line (see Fig. \ref{figthree}).

\subsection{Experimental data}

We tested our model predictions using the open-source dataset of neuropixel recordings from the Allen Institute for Brain Science \cite{siegle2019survey}. We focused our analysis on experiments where drifting gratings were presented at four directions (0\degree, 45\degree, 90\degree, 135\degree) and one temporal frequency (2 Hz). Out of the 54 sessions provided, only 7 sessions had enough trials per behavioral condition to perform our decoding analysis. \com{Neural activity from the visual cortical hierarchy was collected and, specifically: primary visual cortex (V1) in 5 of these 7 sessions, with a median value of 75 neurons per session; lateral visual area (LM): 6 sessions, 47 neurons; anterolateral visual area (AL): 5 sessions, 61 neurons; posteromedial visual area (PM): 6 sessions, 55; anteromedial visual area (AM): 7 sessions, 48 neurons}. We matched the number and duration of trials across condition and orientation and combined trials from the drifting gratings repeat stimulus set, and drifting grating contrast stimulus set. To do this, we combined trials with low-contrast gratings (0.08, 0.1, 0.13, 0.2; see Fig. \ref{figseight}) and trials with high-contrast gratings (0.6, 0.8, 1; see Fig. \ref{figsseven}) into separate trial types to perform the decoding analysis, and analyzed the interval $[-0.25,0.5]$ seconds aligned to stimulus onset. 

For evoked activity, running trials were classified as those where the animal was running faster than 3 cm/s for the first 0.5 seconds of stimulus presentation. During ongoing activity, behavioral periods were broken up into windows of 1 second. Periods of running or rest were classified as such if 10 seconds had elapsed without a behavioral change. Blocks of ongoing activity were sorted and used based on the length of the behavior. Out of the 54 sessions provided, 14 sessions had enough time per behavioral condition (minimum of 2 minutes) to estimate single-cell transfer functions. Only neurons with a mean firing rate during ongoing activity greater than 5Hz were included in the gain analysis (2119 out of 4365 total neurons).

\subsection{Stimulus decoding}

For both the simulations and data, a multi-class decoder was trained to discriminate between four stimuli from single-trial population activity vectors in a given time bin \cite{jezzini2013processing}. To create a timecourse of decoding accuracy, we used a sliding window of 100ms (200ms) in the data (model), which was moved forward in 2ms (20ms) intervals in the data (model). Trials were split into training and test data-sets in a stratified 5-fold cross-validated manner, ensuring equal proportions of trials per orientation in both data-sets. In the model, a leave-2-out cross-validation was performed. To calculate the significance of the decoding accuracy, an iterative shuffle procedure was performed on each fold of the cross-validation. On each shuffle, the training labels were shuffled and the classifer accuracy was predicted on the unshuffled test data-set. This shuffle was performed 100 times to create a shuffle distribution to rank the actual decoding accuracy from the unshuffled decoder against and to determine when the mean decoding accuracy had increased above chance. This time point is what we referred to as the latency of stimulus decoding. To account for the speed of stimulus decoding (the slope of the decoding curve), we defined the $\Delta$-Latency between running and rest as the average time between the two averaged decoding curves from 40\% up to 80\% of the max decoding value at rest.

\subsection{Firing rate distribution match}

To control for increases of firing rate due to locomotion (Fig. \ref{figseven}b), we matched the distributions of population counts across the trials used for decoding in both behavioral conditions. This procedure was done independently for each sliding window of time along the decoding time course. Within each window, the spikes from all neurons were summed to get a population spike count per trial. A log-normal distribution was fit to the population counts across trials for rest and running before the distribution match (Fig \ref{figsfive}a left). We sorted the distributions for rest and running in descending order, randomly removing spikes from trials in the running distribution to match the corresponding trials in the rest distribution (Fig \ref{figsfive}a right). By doing this, we only removed the number of spikes necessary to match the running distribution to rest distribution. For example, trials where the rest distribution had a larger population count, no spikes were removed from either distribution. Given we performed this procedure at the population level rather than per neuron, we checked the change in PSTH between running and rest conditions before and after distribution matching (Fig \ref{figsfive}b). This procedure was also performed on the simulated data (Fig. \ref{figsnine}).

\subsection{Single-cell gain}

To infer the single-cell transfer function in simulations and data, we followed the method originally described in \cite{recanatesi2020metastable} (see also \cite{lim2015inferring,pereira2018attractor} for a trial-averaged version). We estimated the transfer function on ongoing periods when no sensory stimulus was present. Briefly, the transfer function of a neuron was calculated by mapping the quantiles of a standard gaussian distribution of input currents to the quantiles of the empirical firing rate distribution during ongoing periods (Fig. \ref{figthree}d). We then fit this transfer function with a sigmoidal function. The max firing rate of the neuron in the sigmoidal fit was bounded to be no larger than 1.5 times that of the empirical max firing rate, to ensure realistic fits. We defined the gain as the slope at the inflection point of the sigmoid.

\subsection{Single-cell response and selectivity}

We estimated the proportion of neurons that were significantly excited or inhibited by cortical state perturbations in the model (Fig. \ref{figsix}) or locomotion in the data (Fig. \ref{figseven}) during periods of ongoing activity, in the absence of sensory stimuli. In the model, we simulated 40 trials per network, for 10 networks per each value of the perturbation; each trial in the interval $[-0.5,1]$s, with onset of the perturbation at $t=0$ (the perturbation was modeled as a double exponential with rise and decay times $[0.2,1]$, Fig. \ref{figthree}a). In the data, we binned the spike counts in 500ms windows for each neuron after matching sample size between rest and running conditions, and significant difference between the conditions was assessed with a rank-sum test. 

We estimated single neuron selectivity to sensory stimuli in each condition from the average firing rate responses $r_i^a(t)$ of the $i$-th neuron to stimulus $a$ in trial $t$. For each pair of stimuli, selectivity was estimated as
$$
d'(a,b)={\textrm{mean}\left[r(t)^a\right]-\textrm{mean}\left[r(t)^b\right]\over\sqrt{ {1\over2}\left(\textrm{var}[r(t)^a]+\textrm{var}[r(t)^b]\right)}} \ ,
$$
where mean and var are estimated across trials. The d' was then averaged across stimulus pairs.

\begin{figure*}[!t]
\vspace{-0cm}
\includegraphics[width=1\textwidth]{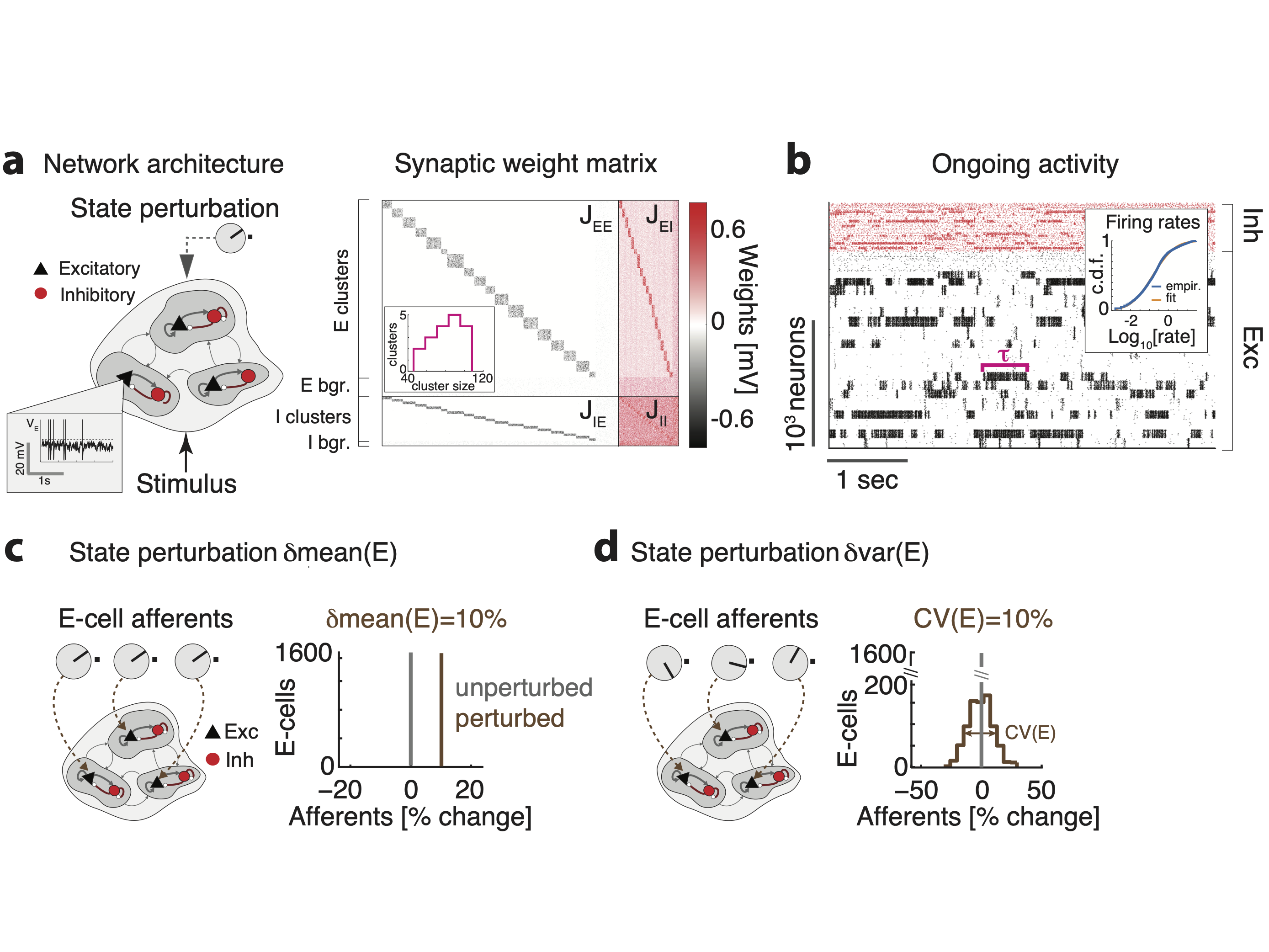}
\vspace{-1cm}
\caption{Biological plausible model of cortical circuit. {\bf a}) Schematics of the network architecture. A recurrent network of E (black triangles) and I (red circles) spiking neurons arranged in clusters is presented sensory stimuli targeting subsets of E clusters, in different cortical states implemented by perturbations. \emph{Inset} shows a membrane potential trace from representative E neuron. {\bf b}) Synaptic couplings $J_{ij}$ for a representative clustered network, highlighting the block diagonal structure of potentiated intra-cluster synaptic weights for both E and I clusters, and the background E and I populations (bgr). Cluster size was heterogeneous (inset). {\bf c}) Representative neural activity during ongoing periods; tick marks represent spike times of E (black) or I (red) neurons. \com{The network dynamics is metastable with clusters transiently activity for periods of duration $\tau$}. Inset: The cumulative distributions of single-cell firing rates (in the representative network are lognormal (blue: empirical data; orange: lognormal fit). {\bf c}) Left: State-changing perturbation affecting the mean of the afferent currents to E populations (knobs represent changes in afferent to three representative E cells compared to the unperturbed state). Right: Histogram of afferent inputs to E-cells in the perturbed state (brown, all neurons receive a 10\% increase in input) with respect to the unperturbed state (grey). {\bf d}) Left: State-changing perturbation affecting the variance of afferent currents to E populations. Right: In the perturbed state (brown), each E-cell's afferent input is constant and sampled from a normal distribution with mean equal to the unperturbed value (grey) and 10\% CV.}
\label{figtwo}
\end{figure*}


\section{Results}

\com{To elucidate the effect of state changes on cortical dynamics,} we modeled the local circuit as a network of recurrently connected excitatory (E) and inhibitory (I) spiking neurons. Both E and I populations were arranged in clusters \cite{Amit1997b,LitwinKumarDoiron2012,mazzucato2015dynamics,schaub2015emergence,mazzucato2019expectation}, where synaptic couplings between neurons in the same cluster were potentiated compared to neurons in different clusters, reflecting the empirical observation of cortical assemblies of functionally correlated neurons \cite[Fig. \ref{figtwo}a;][]{kiani2015natural,song2005highly,perin2011synaptic,lee2016anatomy}. In the absence of external stimulation (ongoing activity), this E-I clustered network generates rich temporal dynamics characterized by metastable activity operating in inhibition stabilized regime, \com{where E and I post-synaptic currents track each other achieving tight balance (Fig. \ref{figsone})}. \com{A heterogeneous distribution of cluster sizes leads to a} lognormal distributions of firing rates (Fig. \ref{figtwo}b). Network activity was characterized by the emergence of the slow timescale of cluster transient activation, with average activation lifetime of $\tau=106\pm35$ ms (hereby referred to as ``cluster timescale," Fig. \ref{figtwo}b), much larger than single neuron time constant \cite[20ms;][]{LitwinKumarDoiron2012,mazzucato2015dynamics}. 

To investigate how changes in cortical state may affect the network dynamics and information processing capabilities, we examined a vast range of state-changing perturbations (Fig. \ref{figtwo}c-d, Table \ref{tablestats}). State changes were implemented as perturbations of the afferent currents to cell-type specific populations, or as perturbations to the synaptic couplings. The first type of state perturbations $\delta$mean(E) affected the mean of the afferent currents to E populations (Fig. \ref{figtwo}c). E.g., a perturbation $\delta$mean(E)=10\% implemented an increase of all input currents to E neurons by 10\% above their unperturbed levels. The  perturbation $\delta$mean(I) affected the mean of the afferent currents to I populations in an analogous way. The second type of state perturbations $\delta$var(E) affected the across-neuron variance of afferents to E populations. Namely, in this perturbed state, the afferent current to each neuron in that population was sampled from a normal distribution with zero mean and fixed variance (Fig. \ref{figtwo}d, \com{measured by the} coefficient of variability CV(E)=var(E)/mean(E) with respect to the unperturbed afferents). This perturbation thus introduced a {\it spatial variance} across neurons in the cell-type specific afferent currents, yet left the mean afferent current into the population unchanged. The state perturbation $\delta$var(I) affected the variance of the afferent currents to I populations analogously. In the third type of state perturbations $\delta$AMPA or $\delta$GABA, we changed the average GABAergic or glutamatergic (AMPA) recurrent synaptic weights compared to their unperturbed values. We chose the range of state perturbations such that the network still retained non-trivial metastable dynamics within the whole range.  \com{We will refer to these state changes of the network as simply perturbations, and should not be confused with the presentation of the stimulus. We first established the effects of perturbations on ongoing network dynamics, and used those insight to explain their effects on stimulus-evoked activity.} 

\com{\subsection{State-dependent regulation of the network emergent timescale}
}

\com{A crucial feature of neural activity in clustered networks is metastable attractor dynamics, characterized by the emergence of a long timescale of cluster activation whereby network itinerant activity explores the large attractor landscape (Fig. \ref{figtwo}b). We first \com{examined} whether perturbations modulated the network's \com{metastable} dynamics and introduced a protocol where perturbations occurred in the absence of sensory stimuli (``ongoing activity'').
}

We found that perturbations strongly modulated the attractor landscape, changing the repertoire of attractors the network activity visited during its itinerant dynamics (Fig. \ref{figthree}a and \ref{figstwo}a). Changes in attractor landscape were perturbation-specific. Perturbations increasing $\delta$mean(E) ($\delta$mean(I)) induced a consistent shift in the repertoire of attractors: larger perturbations led to larger (smaller) numbers of co-active clusters. Surprisingly, perturbations that increased $\delta$var(E) ($\delta$var(I)), led to network configurations with larger (smaller) sets of co-activated clusters. This effect occurred despite the fact that such perturbations did not change the mean afferent input to the network. Perturbations affecting $\delta$AMPA and $\delta$GABA had similar effects to $\delta$mean(E) and $\delta$mean(I), respectively.

\com{We then examined whether perturbations affected the cluster activation timescale. We found that perturbations differentially modulated the average cluster activation timescale $\tau$ during ongoing periods, in the absence of stimuli (Fig. \ref{figthree}b). In particular, increasing $\delta$mean(E), $\delta$var(E), or $\delta$AMPA led to a proportional acceleration of the network metastable activity and shorter $\tau$; while increasing $\delta$mean(I), $\delta$var(I) or $\delta$GABA induced the opposite effect with longer $\tau$. Changes in $\tau$ were congruent with changes in the duration of intervals between consecutive activations of the same cluster (cluster inter-activation intervals, Fig. \ref{figstwo}).} 

\com{
\subsection{Changes in cluster timescale are controlled by gain modulation}

What is the computational mechanism mediating the changes in cluster timescale, induced by the perturbations? We \com{investigated this question using mean field theory, where network attractors, defined by sets of co-activated clusters, are represented as potential wells in an attractor landscape} \cite{Mascaro1999,LitwinKumarDoiron2012,mazzucato2015dynamics,MattiaSanchezVives2012,mazzucato2019expectation}. Let us illustrate this in a simplified network with two clusters (Fig. \ref{figthree}c and \ref{figsthree}). Here, the attractor landscape consists of two potential wells, each well corresponding to a \com{configuration} where one cluster is active and the other is inactive. When the network activity dwells in the attractor represented by the left potential well, it may escape to the right potential well due to internally generated variability. This process will occur with a probability determined by the height $\Delta$ of the barrier separating the two wells: the higher the barrier, the less likely the transition  \cite{hanggi1990reaction,LitwinKumarDoiron2012,MattiaSanchezVives2012,mazzucato2019expectation}. \com{Mean}


\begin{figure*}[!t]
\vspace{-2cm}
\includegraphics[width=1\textwidth]{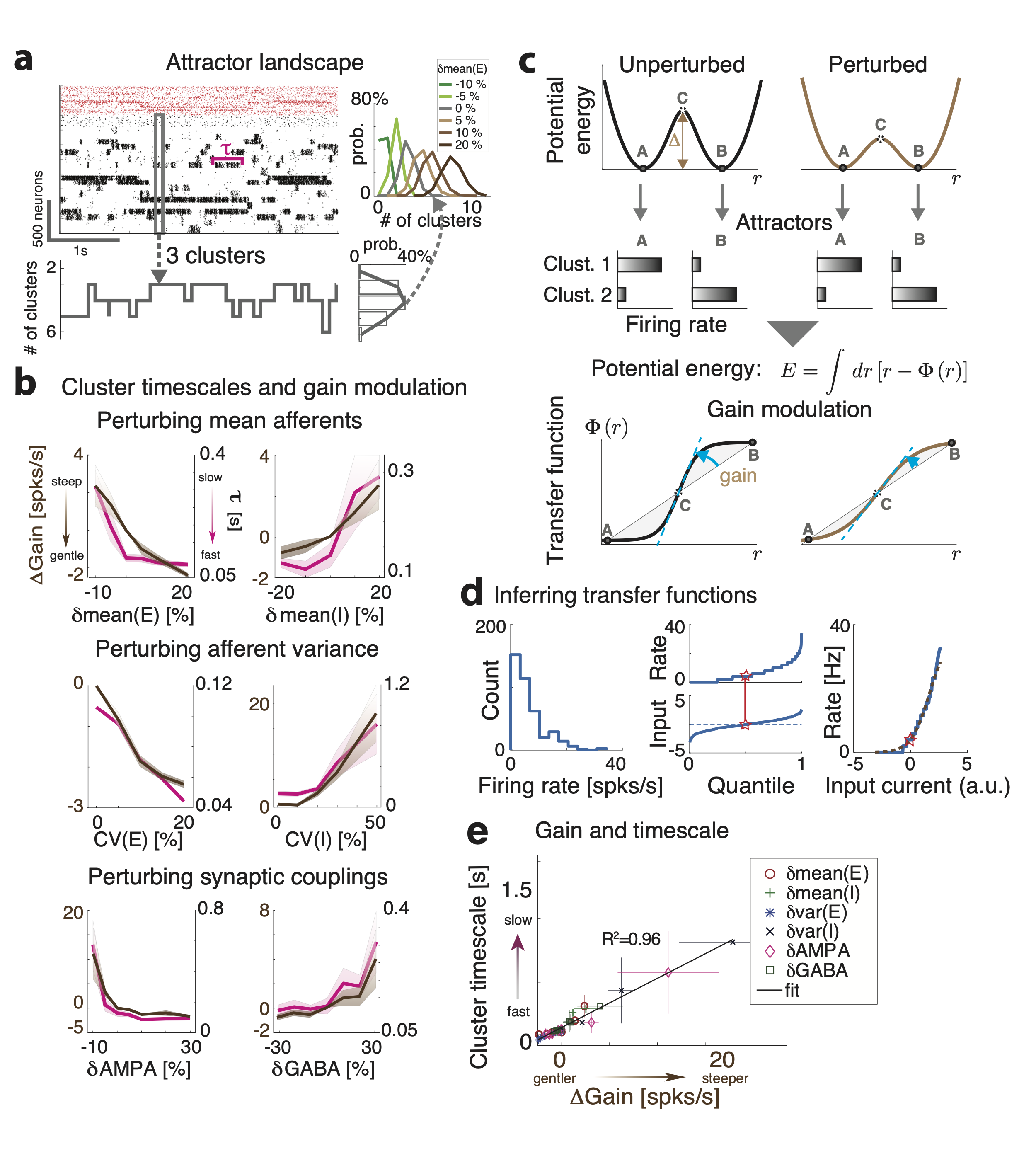}
\vspace{-1cm}
\caption{Linking gain modulation to changes in cluster timescale. {\bf a}) Top left: Clustered network activity during a representative ongoing trial hops among different metastable attractors (grey box: attractor with 3 co-active clusters). Bottom: Number of co-active clusters at each time bin (right: frequency of occurrence of attractors with 2-6 co-active clusters in the representative trial). Right: Perturbations strongly modulate the attractor landscape (color-coded curves: frequency of occurrence of network attractors with different number of co-active clusters, for different values of the representative $\delta$mean(E) perturbation, mean occurrence across 5 sessions). {\bf b}) Perturbations induce consistent changes in the average cluster activation timescale $\tau$ (mean$\pm$S.D. across 5 simulated sessions) and in the single neuron intrinsic gain (estimated as in panel {\bf d}). {\bf c}) Schematic of the effect of perturbations on network dynamics. Dynamics in a two-cluster network is captured by its effective potential energy (top panel). Potential wells represent two attractors where either cluster is active (A and B). Perturbations that shrink the barrier height $\Delta$ separating the attractors induce faster transition rates between attractors and shorter cluster activation lifetime (black and brown: unperturbed and perturbed conditions, respectively). 
}
\label{figthree}
\end{figure*}
{\noindent field theory thus established a relationship between the cluster timescale and the height of the barrier separating the two attractors.} We found that perturbations differentially control the height of the barrier $\Delta$ separating the two attractors (Fig. \ref{figsthree}), {explaining the changes in cluster timescale observed in the simulations (Fig. \ref{figthree}b).}

Since reconstruction of the attractor landscape requires knowledge of the network's structural connectivity, the direct test of the mean field relation between changes in attractor landscape and timescale modulations is challenging. We thus aimed at obtaining an alternative formulation of the underlying neural mechanism only involving quantities directly accessible to experimental observation. Using mean field theory, one can show that the double potential well representing the two attractors can be directly mapped to the effective transfer function of a neural population \cite{Mascaro1999,mazzucato2019expectation, mattia2013heterogeneous}. One can thus establish a direct relationship between changes in the slope (hereby referred to as ``gain") of the intrinsic transfer function estimated during ongoing periods and changes in the barrier height $\Delta$  separating metastable attractors (see Fig. \ref{figthree}c, \ref{figsthree} and Methods). In turn, this implies a direct relationship between gain modulation, induced by the perturbations, and changes in cluster activation timescale. In particular, perturbations inducing steeper gain will increase well depths and barrier heights, and thus increase the cluster timescale, and vice versa. Using mean field theory, we demonstrated a complete classification of the differential effect of all perturbations on barrier heights and gain (Fig. \ref{figsthree}). 

We then proceeded to verify these theoretical predictions, obtained in a simplified two-cluster network, in the high dimensional case of large networks with several clusters using simulations. While barrier heights and the network's attractor landscape can be exactly calculated in the simplified two-cluster network, this task is infeasible in large networks with a large number of clusters where the number of attractors is exponential in the number of clusters. On the other hand, changes in barrier heights $\Delta$ are equivalent to changes in gain, and the latter can be easily estimated from spiking activity (Fig. \ref{figthree}b and \ref{figsthree}). We thus tested whether the relation between gain and timescale held in the high-dimensional case of a network with many clusters. We estimated single-cell transfer functions from their spiking activity during ongoing periods, in the absence of sensory stimuli but in the presence of different perturbations (Fig. \ref{figthree}d, \cite{lim2015inferring,recanatesi2020metastable}). We found that network perturbations strongly } modulated single-cell gain in the absence of stimuli, verifying mean field theory predictions in all cases (Fig. \ref{figthree}d). In particular, we confirmed the direct relationship between gain modulation and cluster timescale modulation: perturbations that decreased (increased) the gain also decreased (increased) cluster timescale (Fig. \ref{figthree}e, $R^2=0.96$). For all perturbations, gain modulations explained the observed changes in cluster timescale.
\blfootnote{\small
\noindent\textit{Legend continued}: Mean field theory provides a relation between potential energy and transfer function (bottom panel), thus linking cluster lifetime to neuronal gain in the absence of stimuli (dashed blue line, gain). {\bf d}): A single-cell transfer function (bottom, empirical data in blue; sigmoidal fit in brown) can be estimated by matching a neuron's firing rate distribution during ongoing periods (top) to a gaussian distribution of input currents (center, quantile plots; red stars denotes matched median values). {\bf e}) Perturbation-induced changes in gain (x-axis: gain change in perturbed minus unperturbed condition, mean$\pm$s.e.m. across 10 networks; color-coded markers represent different perturbations) explain changes in cluster lifetime (y-axis, linear regression, $R^2=0.96$) as predicted by mean field theory (same as in panel {\bf b}).}

\subsection{Controlling information processing speed with perturbations}

\com{
We found that changes in cortical state during ongoing activity, driven by external perturbations, control the circuit's dynamical timescale. The neural mechanism mediating the effects of external perturbations is gain modulation, which controls the timescale of the network switching dynamics. How do such state changes affect the network information processing speed? }

\begin{figure*}[!t]
\includegraphics[width=1\textwidth]{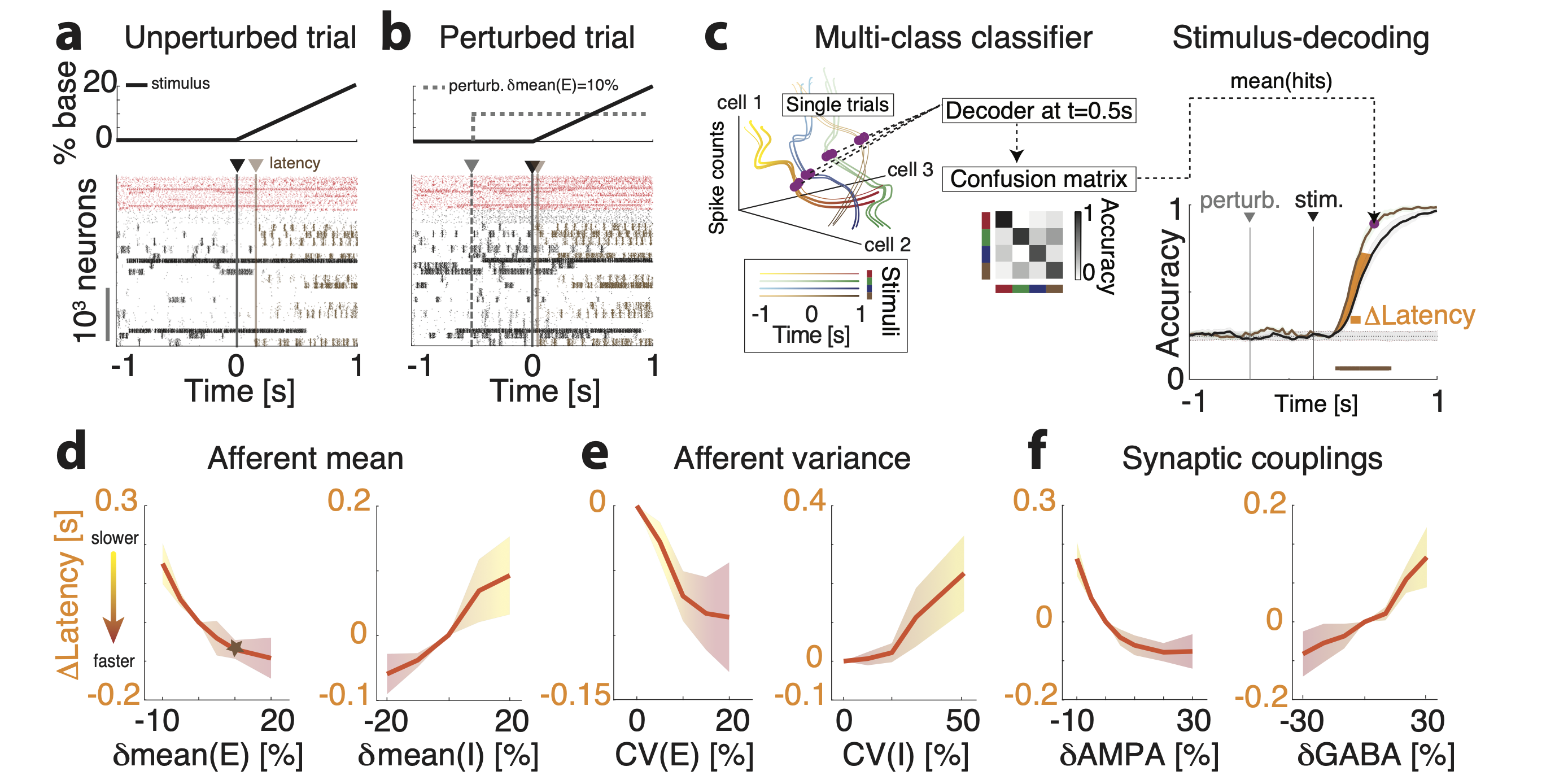}
\vspace{-0cm}
\caption{Perturbations control stimulus-processing speed in the clustered network. {\bf a-b}) Representative trials in the unperturbed ({\bf a}) and perturbed ({\bf b}) conditions; the representative perturbation is an increase in the spatial variance $\delta$var(E) across E neurons. After a ramping stimulus is presented at $t=0$ (black vertical line on raster plot; top panel, stimulus time course), stimulus-selective E-clusters (brown tick marks represent their spiking activity) are activated at a certain latency (brown vertical line). In the perturbed condition ({\bf b}), a perturbation is turned on before stimulus onset (gray-dashed vertical line). The activation latency of stimulus-selective clusters is shorter in the perturbed compared to the unperturbed condition. {\bf c}) Left: schematic of stimulus-decoding analysis. A multi-class classifier is trained to discriminate between the four stimuli from single-trial population activity vectors in a given time bin (curves represent the time course of population activity in single trials, color-coded for 4 stimuli; the purple circle highlights a given time bin along the trajectories), yielding a cross-validated confusion matrix for the decoding accuracy at that bin (central panel). Right: Average time course of the stimulus-decoding accuracy in the unperturbed (black) and perturbed (brown) conditions (horizontal brown: significant difference between conditions, $p<0.05$ with multiple bin correction). {\bf d-f}: Difference in stimulus decoding latency in the perturbed minus the unperturbed conditions (average difference between decoding accuracy time courses in the [40\%,80\%] range, yellow interval of {\bf c}; mean$\pm$S.D. across 10 networks) for the six state-changing perturbations (see Methods and main text for details; the brown star represents the perturbation in {\bf b}-{\bf c}).}
\label{figfour}
\end{figure*}

To investigate the effect of state perturbations on the network's information-processing, we compared stimulus-evoked activity by presenting stimuli in an unperturbed and a perturbed condition. In unperturbed trials (Fig. \ref{figfour}a), we presented one of four sensory stimuli, modeled as depolarizing currents targeting a subset of stimulus-selective E neurons \com{with linearly ramping time course}. Stimulus selectivities were mixed and random, all clusters having equal probability of being stimulus-selective. In perturbed trials (Fig. \ref{figfour}b), in addition to the same sensory stimuli, we included a state perturbation, which was turned on before the stimulus and was active until the end of stimulus presentation. 
We investigated and classified the effect of several state changes implemented by perturbations affecting either the mean or variance of cell-type specific afferents to E or I populations, and the synaptic couplings. State perturbations were identical in all trials of the perturbed condition for each type; namely, they did not convey any information about the stimuli.

We assessed how much information about the stimuli was encoded in the population spike trains at each moment using a multiclass classifier (with four class labels corresponding to the four presented stimuli, Fig. \ref{figfour}c). In the unperturbed condition, the time course of the cross-validated decoding accuracy, averaged across stimuli, was significantly above chance after $0.21+/-0.02$ seconds (mean$\pm$s.e.m. across 10 simulated networks, black curve in Fig. \ref{figfour}c) and reached perfect accuracy after a second. In the perturbed condition, stimulus identity was decoded at chance level in the period after the onset of the state perturbation but before stimulus presentation (Fig.~\ref{figfour}c), consistent with the fact that the state perturbation did not convey information about the stimuli. We found that state perturbations significantly modulated the network information processing speed. We quantified this modulation as the average latency to reach a decoding accuracy between 40\% and 80\% (Fig. \ref{figfour}c, yellow area), and found that state perturbations differentially affected processing speed. 

State perturbations had opposite effects depending on which cell-type specific populations they targeted. Increasing $\delta$mean(E) monotonically improved network performance (Fig. \ref{figfour}d, left panel): in particular, positive perturbations induced an anticipation of stimulus-coding (shorter latency), while negative ones led to longer latency and slower coding. The opposite effect was achieved when increasing $\delta$mean(I), which slowed down processing speed (Fig. \ref{figfour}d, right panel). State perturbations that changed the spatial variance of the afferent currents had counterintuitive effects (Fig. \ref{figfour}e). We measured the strength of these perturbations via their coefficient of variability $CV(\alpha)=\sigma_\alpha/\mu_\alpha$, for $\alpha=E,I$, where $\sigma$ and $\mu$ are the standard deviation and mean of the across-neuron distribution of afferent currents. Perturbations $\delta$var(E) that increased $CV(E)$ led to faster processing speed. The opposite effect was achieved with perturbations $\delta$var(I) inducing a spatial variance across afferents to I neurons, which slowed down stimulus-processing speed  (Fig. \ref{figfour}e). Perturbations $\delta$AMPA which increased the glutamatergic synaptic weights improved performance proportionally to the perturbation. The opposite effect was achieved by perturbations $\delta$GABA that increased the GABAergic synaptic weights, which monotonically decreased network processing speed  (Fig. \ref{figfour}f).
We thus concluded that afferent current perturbations differentially modulated the speed at which network activity encoded information about incoming sensory inputs. Such modulations exhibited a rich dynamical repertoire (Table \ref{tablestats}).

\com{\subsection{Gain modulation regulates the network information processing speed}}

\com{Our mean field framework demonstrates a direct relationship between the effects of perturbations on the network information processing speed and its effects on the cluster timescale (Fig. \ref{figthree}). In our simplified network with two clusters, stimulus presentation induces an asymmetry in the well depths, where the attractor B corresponding to the activated stimulus-selective cluster has a deeper well, compared to the attractor A where the stimulus-selective cluster is inactive. Upon stimulus presentation, the network ongoing state will be biased to transition towards the stimulus-selective attractor B with a transition rate determined by the barrier height separating A to B. Because external perturbations regulate the height of such barrier via gain modulation, they control in turn the latency of activation of the stimulus-selective cluster. We thus aimed at testing the prediction of our theory: that the perturbations modulate stimulus coding latency in the same way as they modulate cluster timescales during ongoing periods; and, as a consequence, that these changes in stimulus coding latency can be predicted by intrinsic gain modulation. Specifically, our theory predicts that perturbation driving a decrease (increase) in intrinsic gain during ongoing periods will induce a faster (slower) encoding of the stimulus.}

\begin{figure*}[!t]
\vspace{-1cm}
\includegraphics[width=1\textwidth]{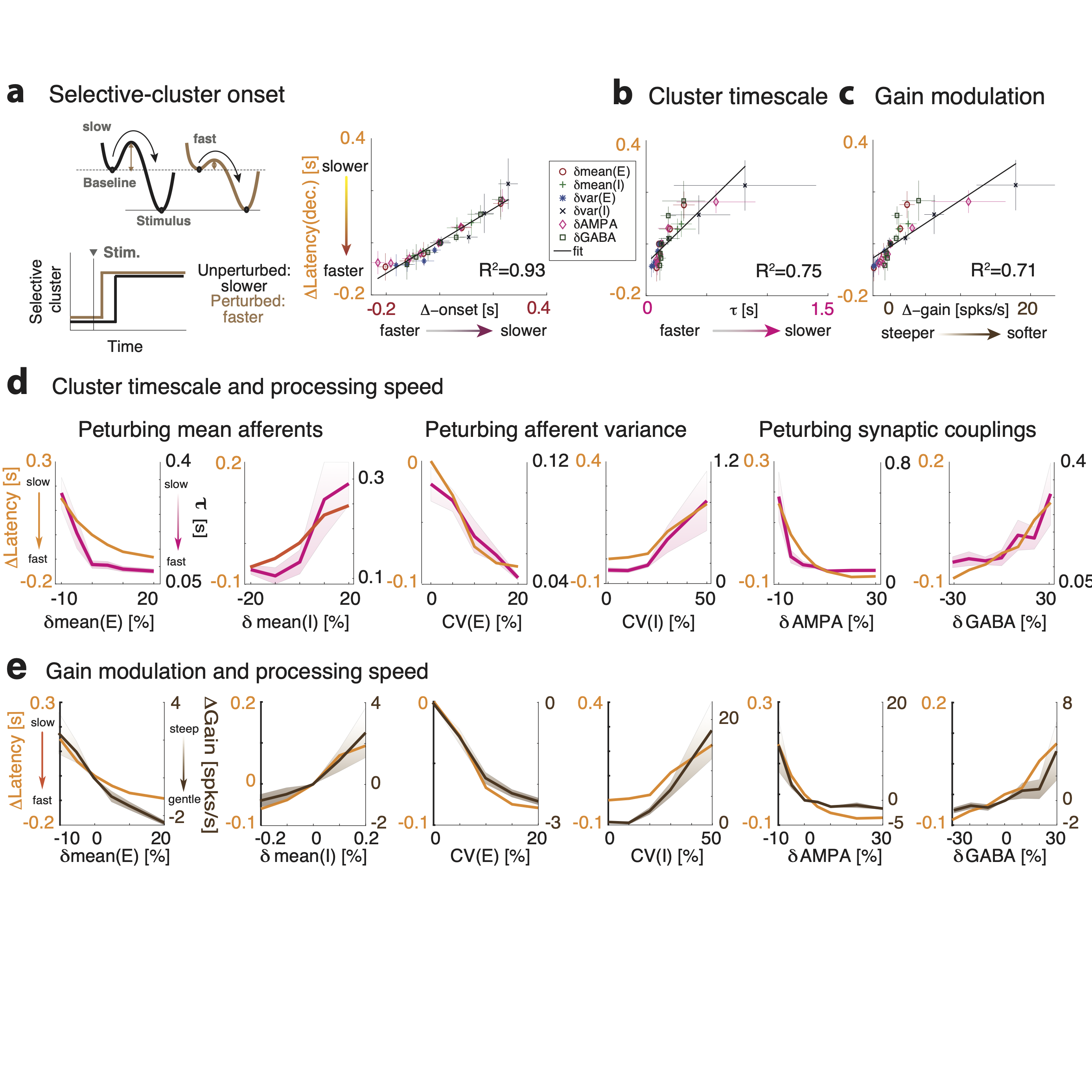}
\vspace{-3cm}
\caption{\com{Linking gain modulation to changes in processing speed. {\bf a}) Left: Schematic of the effect of an accelerating perturbation on stimulus encoding during evoked activity (same notations as in \ref{figthree}c). A shrinking barrier from non-selective to selective attractors drives a faster activation of stimulus-selective cluster after stimulus presentation. Right: Changes in stimulus processing speed (y-axis: latency of stimulus decoding from \ref{figfour}d-f) are predicted by changes in activation latency of stimulus selective clusters (x-axis: mean$\pm$S.D. across 10 simulated sessions; linear regression, $R^2=0.93$); {\bf b, d}) by changes in cluster timescale (same values as \ref{figthree}b; $R^2=0.93$); {\bf c, e}) and by changes in single-cell intrinsic gain (same values as \ref{figthree}b; $R^2=0.71$).}}
\label{figfive}
\end{figure*}

\com{We thus proceeded to test the relationship between perturbations effects on cluster timescales, gain modulation, and information processing speed. In the representative trial where the same stimulus was presented in the absence (Fig. \ref{figfour}a) or in the presence (Fig. \ref{figfour}b) of the perturbation $\delta$mean(E)$=10\%$, we found that stimulus-selective clusters (highlighted in brown) had a faster activation latency in response to the stimulus in the perturbed condition compared to the unperturbed one. A systematic analysis confirmed this mechanism showing that, for all perturbations, the activation latency of stimulus-selective clusters was a strong predictor of the change in decoding latency (Fig. \ref{figfive}a right panel, $R^2=0.93$). Moreover, we found that the perturbation-induced changes of the cluster timescale $\tau$ during ongoing periods predicted the effect of the perturbation on stimulus-processing latency during evoked periods (Fig. \ref{figfive}b,d). Specifically, perturbations inducing faster $\tau$ during ongoing periods, in turn accelerated stimulus coding; and vice versa for perturbations inducing slower $\tau$. 

We then tested whether perturbation-induced gain modulations during ongoing periods explained the changes in stimulus-processing speed during evoked periods, and found that the theoretical prediction was borne out in the simulations (Fig. \ref{figfive}c,e). Let us summarize the conclusion of our theoretical analyses. Motivated by mean field theory linking gain modulation to changes in transition rates between attractors, we found that gain modulation controls the cluster timescale during ongoing periods, and, in turn, regulates the onset latency of stimulus-selective clusters upon stimulus presentation. Changes in onset latency of stimulus-selective clusters explained changes in stimulus-coding latency. We thus linked gain modulation to changes in stimulus-processing speed (Fig. \ref{figfive}, Table \ref{tablestats}). 
}

\begin{table}
\centering
\begin{tabular}{l|ccccc}
&Latency & Activity & Response & $\tau$ & Gain \\\hline
$\delta$mean(E)$\nearrow$ & $\searrow$ & E[$\nearrow$], I[$\nearrow$] & E[$\nearrow$], I[$\nearrow$] &$\searrow$ & $\searrow$\\
$\delta$mean(I)$\nearrow$ & $\nearrow$ & E[$\searrow$], I[$\searrow$] & E[$\searrow$], I[mixed] & $\nearrow$ & $\nearrow$\\
$\delta$var(E)$\nearrow$ & $\searrow$ & E[$\nearrow$], I[$\nearrow$] & E[mixed], I[mixed] & $\searrow$ & $\searrow$\\
$\delta$var(I)$\nearrow$ & $\nearrow$ & E[$\searrow$], I[$=$] & E[$\searrow$], I[$\searrow$] & $\nearrow$ & $\nearrow$\\
$\delta$AMPA$\nearrow$ & $\searrow$ & E[$\nearrow$], I[$\nearrow$] & E[$\nearrow$], I[$\nearrow$] & $\searrow$ & $\searrow$\\
$\delta$GABA$\nearrow$ & $\nearrow$ & E[$\searrow$], I[$\searrow$] & E[$\searrow$], I[mixed] & $\nearrow$ & $\nearrow$\\
Locomotion & $\searrow$ & E[$\nearrow$], I[$\nearrow$] & E[mixed], I[mixed] & $\searrow$ & $\searrow$\\
\end{tabular}
\caption{Classification of state-changing perturbations. Effect of on neural activity of an increasing $(\nearrow)$ state-changing perturbation: latency of stimulus decoding ('Latency', Fig. \ref{figtwo}d); average firing rate modulation ('Activity') and response to perturbations ('Response', proportion of cells with significant responses) of E and I cells in the absence of stimuli (Fig. \ref{figthree}); cluster activation timescale ('$\tau$', Fig. \ref{figfour}b); single-cell intrinsic gain modulation at rest ('Gain', Fig. \ref{figfive}e). $\nearrow,\searrow,=$ represent increase, decrease, and no change, respectively. 'Mixed' responses refer to similar proportions of excited and inhibited cells. The effect of locomotion is consistent with a perturbation increasing $\delta$var(E).}
\label{tablestats}
\end{table}

\subsection{Physiological responses to perturbations}

\com{Our results show that cortical processing speed can be accelerated or slowed down via external perturbations. We found that different types of perturbations may induce similar effects on processing speed: a dynamical acceleration may be obtained by either increasing the mean or the variance of the external input to E neurons, or either decreasing the mean or the variance of external inputs to I neurons. A dynamical deceleration may be obtained by the opposite perturbations. In order to devise an experimental test of our theory to dissect the specific effects of each type of perturbation, we then examined \com{the single-cell responses to perturbations}. By combining single-cell responses with dynamical effects, we will be able to isolate the effects of each perturbation.}

\begin{figure*}[!t]
\vspace{-1cm}
\includegraphics[width=1\textwidth]{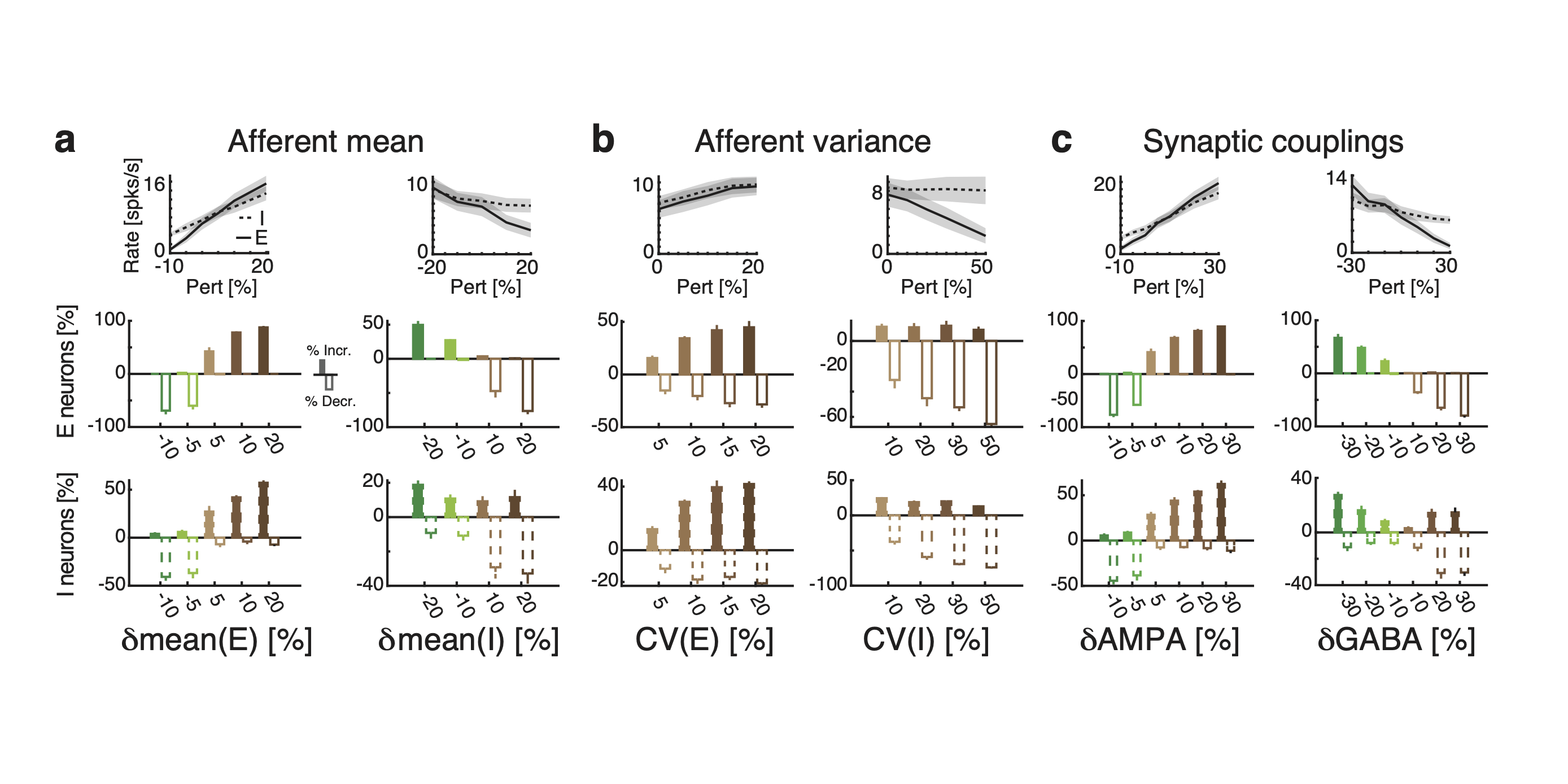}
\vspace{-1cm}
\caption{Single-cell responses to perturbations. Top: Average firing rate change across E (full) and I (dashed) populations in response to each state-changing perturbations (mean$\pm$S.D. across 10 simulated networks). Histograms: Average fractions of E (top row) and I (bottom row) neurons whose firing rate significantly increase (positive bars) or decrease (negative bars) in responses to the perturbations (single-cell significant response was based on a t-test of the baseline vs. perturbation evoked activity, $p<0.05$).}
\label{figsix}
\end{figure*}

We characterized single-cell responses to perturbations during ongoing periods, in the absence of sensory stimuli (Fig. \ref{figsix}, \ref{figsfour}). We found that perturbations differentially affected neuronal responses in a cell-type specific way. Perturbations changed the average population firing rates, and led to complex patterns of response across E and I populations (Fig. \ref{figsix}). Specifically, perturbations increasing $\delta$mean(E) induced higher firing rates and induced proportionally excited responses in both E and I populations. On the other hand, perturbations that increased $\delta$mean(I) led to a decrease in both E and I average firing rates (Fig. \ref{figsix}). This paradoxical effect \cite{tsodyks1997paradoxical} revealed that the network operates in the inhibition stabilized regime (Fig. \ref{figsone}). When increasing the inhibitory current beyond $\delta$mean(I)=$50\%$, the network reached a reversal point where the E population activity became silent and the I population rebounded, starting to increase their firing rates again (Fig. \ref{figsone}). 

Perturbations increasing the variance $\delta$var(E) and $\delta$var(I) led to surprising effects (Fig. \ref{figsix}, \ref{figsfour}). Increasing $\delta$var(E) induced higher firing rates in both E and I populations, despite leaving the mean afferents unchanged; moreover, it led to mixed responses at the single cell level, with a prevalence of excited responses in both E and I populations. We will see below that this set of responses is consistent with locomotion-induced effects in the visual cortical hierarchy. Increasing $\delta$var(I) left firing rates of I populations unchanged but led to a decrease of E population firing rates. This perturbation also induced mixed responses at the single cell level, with a prevalence of inhibited responses in both populations. Finally, perturbations $\delta$AMPA and $\delta$GABA led to responses similar to those found when driving the mean E- or I-afferents, respectively.

\com{Our theory thus suggests that it is possible to identify a specific perturbation by combining all its effects, including gain modulation, changes in stimulus-processing speed and single-cell physiological responses (Table \ref{tablestats})
.}

\subsection{{Changes in single-cell responses cannot explain the effects of perturbations on evoked activity}}

\com{Although we found that gain modulation captures the effects of perturbations on network activity, we investigate whether alternative explanations were also possible, in terms of traditional measures of stimulus responsiveness and selectivity.}

We found that perturbations strongly affected the peak of single-cell responses to stimuli compared to baseline ($\Delta$PSTH, Fig. \ref{figsfour}b), as well as single-cell selectivity to stimuli with significant changes in their d' (Fig. \ref{figsfour}c). 
We then tested whether perturbation-induced changes in stimulus responses or selectivity could explain the observed changes in stimulus-processing speed.
We first hypothesized that, if the response increase induced by the perturbation were larger for stimulus-selective compared to nonselective neurons (i.e., if $\Delta$PSTH(sel)>$\Delta$PSTH(nonsel)), then a perturbation increasing stimulus-responses could lead to faster stimulus-processing speed. Likewise, we hypothesized that faster stimulus-processing speed may be induced by perturbations improving single-cell selectivity (d') to stimuli. Surprisingly, we found a complex relation between changes in single-cell responsiveness and selectivity to stimuli, induced by the perturbations, and modulation of stimulus-processing speed (Fig. \ref{figsfour}). For perturbations targeting I populations ($\delta$mean(I), $\delta$var(I), and $\delta$GABA) changes in responsiveness and selectivity were consistent with changes in processing speed ($R^2=0.92,0.70$ for responsiveness and selectivity, respectively). However, for perturbations targeting E populations ($\delta$mean(E), $\delta$var(E), and $\delta$AMPA) changes in responsiveness and selectivity were not consistent with changes in processing speed ($R^2=0.05,0.02$ for responsiveness and selectivity, respectively). Strikingly, in the case of the perturbation $\delta$var(E), processing speed increased with larger perturbations even though responses and selectivity increasingly degraded. In the case of the perturbation $\delta$mean(E) and $\delta$AMPA, network performance likewise increased but single-cell metrics where non-monotonic in the value of the perturbation (Fig. \ref{figsfour}e, f). Because changes in single-cell stimulus properties were only consistent with changes in processing speed for some perturbations ($\delta$mean(I), $\delta$var(I), and $\delta$GABA), but inconsistent for other perturbations, we thus conclude that they could not represent an alternative mechanism underlying the observed effects of perturbations.

\subsection{Locomotion decreases single-cell gain and accelerates visual processing speed}

\begin{figure*}[!t]
\vspace{-1cm}
\includegraphics[width=1\textwidth]{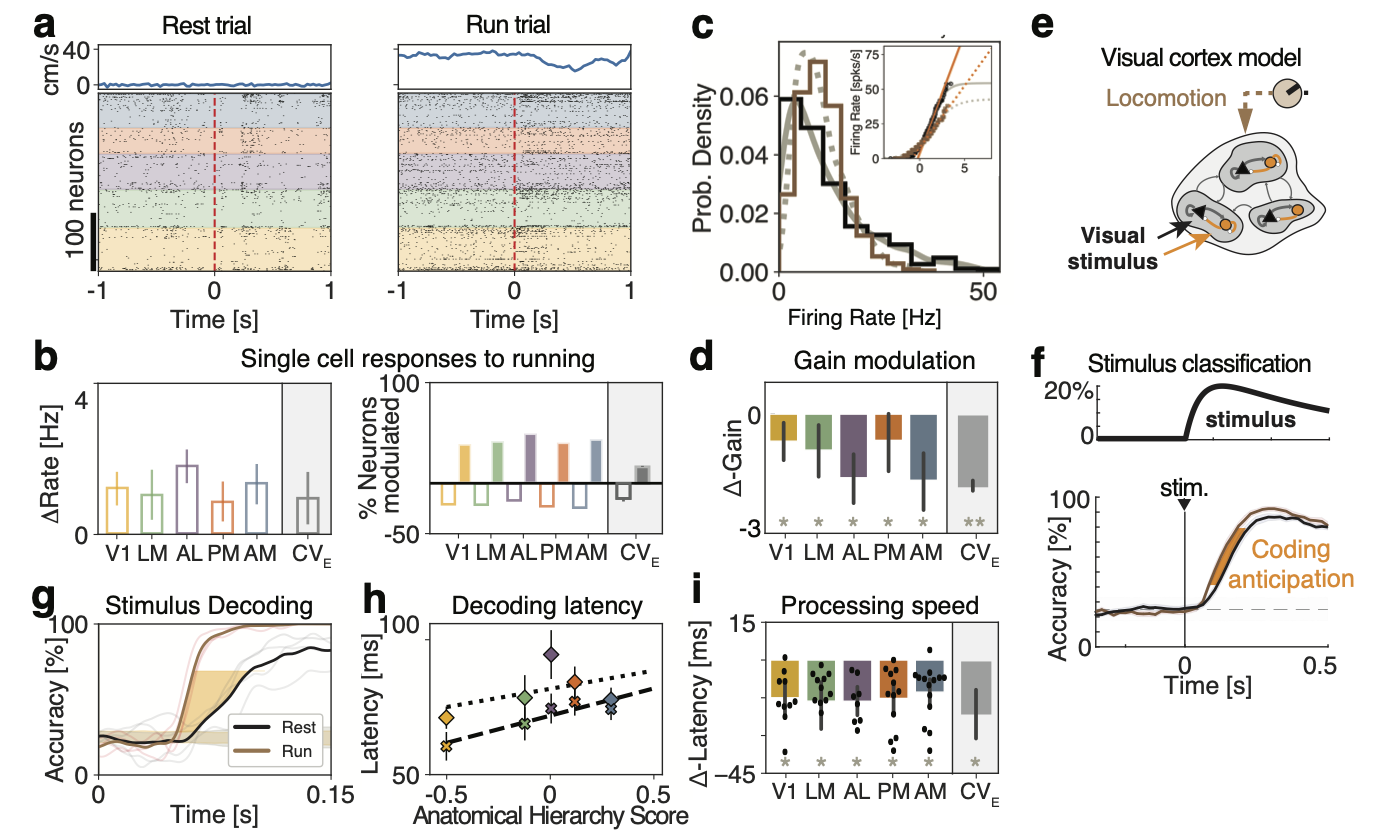}
\vspace{-0.5cm}
\caption{\com{Locomotion effects on visual processing are mediated by gain modulation}. {\bf a}) Representative raster plots from five cortical visual areas (color-coded) with population spiking activity during passive presentation of drifting gratings (dashed red line represents stimulus onset) during periods of running (right, running speed in top panels) rest (left).{\bf b}) Left: Firing rate modulation induced by running per area (colors) and in the model (gray, CV(E)=5\%), averaged across all periods of ongoing activity. Right: Fraction of neurons by area (colors) and in the model (gray, CV(E)=5\%) with significantly excited (positive bars) and inhibited (negative bars) responses to bouts of running (rank-sum test, $*=p < 0.005$). {\bf c}): A representative single-cell distribution of firing rates for rest (blue) and running (red) conditions. The overlaid distributions of firing rates are obtained by passing a standard normal distribution through the sigmoidal transfer function fit shown in the inset for rest (full gray line) and running (dashed gray line). The gain for each behavioral condition (orange lines) was estimated as the slope of the sigmoidal transfer function fit at the inflection point (see Methods). {\bf d}): Single-cell gain modulation ($\Delta$gain=gain(running)-gain(rest)) by area (colors) and in the model (gray, CV(E)=5\%) across all neurons during ongoing periods (bars show 95\% confidence interval; rank-sum test $*=p < 0.005$). {\bf e}) Time course of the mean stimulus-decoding accuracy across orientations during running and rest using neurons from V1 as predictors shows the anticipation of stimulus coding in the running condition (single sessions and session average, thin and thick lines, respectively; see Methods). {\bf f} Decoding latency (first bin above chance decoding regions in {\emph e}) slows down along the anatomical hierarchy (x-axis: anatomical hierarchy score from \cite{siegle2019survey}). Dotted (dashed) line with diamond ("x") symbols show the latency during rest (running). {\bf g}) Difference in processing speed between running and resting (average latency of decoding accuracy between 40\% and 80\%, yellow area in panel e) reveals running-induced coding acceleration in all areas (colors) and in the model (gray, CV(E)=5\%). t-test, $*=p < 0.01$. }
\label{figseven}
\end{figure*}

Our theory predicts a link between gain modulations measured during ongoing periods and changes in stimulus-processing speed \com{during evoked periods}. We sought to experimentally test this prediction in freely running mice using electrophysiological recordings from \com{the visual hierarchy including the} primary visual cortex (V1) and 4 higher cortical visual areas LM, AL, PM, AM \cite[open-source neuropixels dataset available from the Allen Institute,][]{siegle2019survey}. 
We interpreted periods where the animal was resting as akin to the ``unperturbed'' condition in our model, and periods where the animal was running as the ``perturbed'' condition  (Fig. \ref{figseven}a in the data). \com{We thus set out to test our theory in the following three steps: i) in each area, we estimated the effect of locomotion on single-cell responses and on intrinsic gain  during ongoing periods, ii) based on these changes, we built a biologically plausible model of cortical circuits processing visual stimuli and predicted whether locomotion would accelerate or slow down visually-evoked responses; iii) we tested the prediction in each area with a decoding analysis of visually-evoked population activity.}

During periods of ongoing activity (in the absence of visual stimuli), we found that locomotion induced an overall increase in firing rate across all visual cortical areas (Fig. \ref{figseven}b left), in agreement with previous studies \cite{niell2010modulation,stringer2019high,dipoppa2018vision}. \com{Although, we found that locomotion led to complex responses inducing mixed excited and inhibited responses across neurons (Fig. \ref{figseven}b right)}, as previously reported \cite{dipoppa2018vision}. We then estimated the single-cell transfer functions from spiking activity during ongoing periods both when the animal was at rest and in motion  (Fig. \ref{figseven}c). We found that locomotion strongly modulated the single-cell gain in the absence of stimuli \com{in all visual cortical areas (Fig. \ref{figseven}d). Specifically, we found that locomotion on average decreased the single-cell gain.}

{Our theory predicts that, in all visual cortical areas}, the locomotion-induced increase in firing rates, the mixed excited and inhibited neural responses and the decrease in intrinsic gain are consistent with a state-changing perturbation mediated by an increase in the variance of the input currents to E neurons ($\delta$var(E), Table \ref{tablestats}). According to our theory, the decrease in gain leads to an acceleration of stimulus-processing speed \com{in all visual cortical areas.}

\com{We aimed at refining the model predictions on the locomotion effects on V1 and the visual cortical hierarchy by introducing a biologically plausible stimulation protocol in our spiking network. Following experimental evidence on anatomical connectivity in the visual pathway \cite{zhuang2013layer,miska2018sensory,kloc2014target,khan2018distinct}, we then modeled incoming visual stimuli as a transient increase in the input currents to both E and I neurons (Fig. \ref{figseven}e). We then modeled the effect of locomotion as an external perturbation inducing an increase in the variance of the inputs to E neurons $\delta$var(E), capturing the observed empirical effects of locomotion on ongoing periods in terms of gain decrease and mixed single-cell responses. In this  model of visual processing, we found that locomotion accelerated visual processing speed during evoked period by $21\pm9$ms on average (mean$\pm$S.D. across 10 sessions, Fig. \ref{figseven}f). We thus set out to test this prediction in the empirical data.
}

Previous studies have observed an improvement in peak decoding performance during locomotion \cite{dadarlat2017locomotion}, but changes in decoding latency have not been investigated. To probe the speed and accuracy of visual responses in perturbed and unperturbed conditions, we performed a cross-validated classification analysis to assess the amount of information regarding the orientation of drifting grating stimuli present in population spiking activity along the visual cortical hierarchy. Crucially, because decoding accuracy depends on sample size, we equalized number of trials between resting and running conditions. We found that trials in which the animal was running revealed both an increase in peak decoding accuracy and an anticipation of stimulus coding (shorter latency) as compared to trials where the animal was stationary (Fig. \ref{figseven}g), consistently across the whole visual hierarchy (Fig. \ref{figseven}i). Furthermore, the time to reach significant decoding for each \com{cortical} area followed the anatomical hierarchy score in both unperturbed and perturbed conditions, consistent with the idea that information about the visual stimulus travels up a visual hierarchy in a feed-forward fashion \cite[Fig. \ref{figseven}h,][]{siegle2019survey}. 

Given that locomotion induced an increase in firing rates in all \com{cortical} areas (Fig. \ref{figseven}b), we then examined the extent to which the observed effects of locomotion (increased peak accuracy and anticipation) were merely due to the increase in firing rates. We thus matched the distribution of firing rates between running and resting (see methods and  Fig. \ref{figsfive}). We found that after rate matching the change in peak decoding accuracy decreased significantly (Fig. \ref{figssix}-\ref{figseight}). Crucially, the anticipation of stimulus processing speed induced by locomotion was still present in the rate-matched condition (Fig. \ref{figssix}), confirming that it was independent of changes in firing rates. The same effect was preserved in the rate-matched model simulations as well (Fig. \ref{figsnine}).
We thus concluded that the anticipation of visual processing speed induced by locomotion is consistent with a mechanism whereby locomotion decreases single-cell gain via an increase in the afferent variance $\delta$var(E) as predicted by our theory (Table \ref{tablestats}).

\section{Discussion}

Cortical circuits flexibly adapt their information processing capabilities to changes in environmental demands and internal state. Empirical evidence suggests that these state-dependent modulations may occur already in the sensory cortex where they may be induced by top-down pathways or neuromodulation. Here, we presented a mechanistic theory explaining how stimulus-processing speed can be \com{regulated} in a state-dependent manner via gain modulation, induced by transient changes in the afferent currents or in the strength of synaptic transmission. 

Our theory entails a recurrent spiking network where excitatory and inhibitory neurons are arranged in clusters, generating metastable activity in the form of transient activation of subsets of clusters. We showed that gain modulation controls the timescale of metastable activity and thus the network's information-processing speed and reaction times \com{upon stimulus presentation}. Specifically, our theory predicted that perturbations that decrease (increase) the intrinsic single-cell gain during ongoing periods accelerate (slow down) the latency of stimulus responses. 

We tested this  prediction by examining the effect of locomotion on visual processing in freely running mice. We found that locomotion reduced the intrinsic single-cell gain during ongoing periods, thus accelerating stimulus-coding speed across the visual cortical hierarchy. Our theory suggests that the observed effects of locomotion are consistent with a perturbation that increases the spatial variance of the afferent currents to the local excitatory population. These results establish a new theory of state-dependent adaptation of cortical responses via gain modulation, unifying the effect of different pathways under a shared computational mechanism.

\subsection{Metastable activity in cortical circuits}

The crucial dynamical feature of our model is its metastable activity, whereby single-trial ensemble spike trains unfold through sequences of metastable states. State are long-lasting, with abrupt transitions between consecutive states. Metastable activity has been ubiquitously observed in a variety of cortical and subcortical areas, across species and tasks \cite{Abeles1995a,Jones2007,engel2016selective,PonceAlvarez2012,maboudi2018uncovering,rich2016decoding,sadacca2016behavioral,taghia2018uncovering,deco2019awakening}. Metastable activity can be used to predict behavior and was implicated as a neural substrate of cognitive function, such as attention \cite{engel2016selective}, expectation \cite{mazzucato2019expectation}, and decision making \cite{rich2016decoding,taghia2018uncovering,recanatesi2020metastable}. Metastable activity was observed also during ongoing periods, in the absence of sensory stimulation, suggesting that it may be an intrinsic dynamical regime of cortical circuits \cite{mazzucato2015dynamics,engel2016selective}. Here, we showed how cortical circuits can flexibly adjust their performance and information-processing speed via modulations of their metastable dynamics. 

Metastable activity may naturally arise in circuits where multiple stable states, or attractors, are destabilized by external perturbations \cite{MillerKatz2010} or intrinsically generated variability \cite{DecoHugues2012,LitwinKumarDoiron2012,mazzucato2015dynamics,schaub2015emergence,mazzucato2019expectation,rostami2020spiking,recanatesi2020metastable}. Biologically plausible models of metastable dynamics have been proposed in terms of recurrent spiking networks where neurons are arranged in clusters, reflecting the empirically observed  assemblies of functionally correlated neurons \cite{kiani2015natural,song2005highly,perin2011synaptic,lee2016anatomy}. Clustered network models of metastable dynamics provide a parsimonious explanation of several physiological observations such as stimulus-induced reductions of trial-to-trial variability \cite{DecoHugues2012,LitwinKumarDoiron2014,mazzucato2015dynamics,rostami2020spiking,Churchland2010}, of firing rate multistability \cite{mazzucato2015dynamics}, and of neural dimensionality \cite{mazzucato2016stimuli}. \com{Compared to previous models of metastable dynamics, our results extend the biological plausibility of clustered networks in several aspects. The introduction of pairs of E and I clusters induces a tight balance where the E and I contributions to the postsynaptic currents of each neuron closely track each other with opposite signs (Fig. \ref{figsone}), as observed experimentally in cortical circuits \cite{okun2008instantaneous}. Moreover, we showed that these networks operate in the inhibition stabilized regime (Fig. \ref{figsone}), which is believed to be the operational regime of cortical circuits \cite{ozeki2009inhibitory,sanzeni2019inhibition,moore2018rapid}. We showed that a heterogeneous distribution of cluster sizes naturally give rise to lognormal distribution of firing rates (Fig. \ref{figtwo}b inset), as observed in cortical circuits \cite{shafi2007variability,hromadka2008sparse,o2010neural,roxin2011distribution}. We then generalize the results in \cite{mazzucato2019expectation} to establish gain modulation as the general mechanism controlling that state-dependent changes in processing speed in recurrent circuits with metastable dynamics.  This class of models thus provide a biologically plausible, mechanistic link between connectivity, dynamics, and information-processing.}

\subsection{Linking metastable activity to flexible cognitive function via gain modulation}

Recent studies have shown that cortical circuits may implement a variety of flexible cognitive computations by modulating the timescale of their intrinsic metastable dynamics \cite{mattia2013heterogeneous,mazzucato2019expectation,engel2016selective,rich2016decoding,deco2019awakening}. Our results establish a comprehensive framework to investigate the extent of this hypothesis. We propose that gain modulation is the neural mechanism underlying flexible state-dependent cortical computation. Specifically, we showed that gain modulation controls the timescale of metastable dynamics, which, in turn, determines the network's information-processing speed.

\com{Our theoretical framework to link gain modulation to changes in potential barrier heights is based on the effective mean field theory following \cite{Mascaro1999,mattia2013heterogeneous,mazzucato2019expectation}, which we used to reduce a multidimensional system to obtain an effective potential describing a single population. Although this approach is exact in the case of networks with symmetric connectivity, it represents only an approximation to the full network dynamics in the case of networks with asymmetric couplings such as the ones considered in this study \cite{rodriguez2020climbing}. It would be interesting to extend our results to an exact framework by estimating the network Lyapunov function \cite{yan2013nonequilibrium}.  
}

\subsection{Alternative models of gain modulation}

Previous studies have suggested gain modulation as a mechanism to sharpen single-cell tuning curves without affecting selectivity \cite{cardin2008cellular,haider2009rapid}, potentially mediating attention \cite{mcadams1999effects,treue1999feature,rabinowitz2015attention}. In those studies, gain modulation was defined as change in the single-neuron response function to stimuli of increasing contrast. Here, we have taken a different approach and defined gain as the slope of the intrinsic neuronal current-to-rate function during ongoing periods \cite[i.e., in the absence of stimuli, see also][]{chance2002gain,haider2009rapid,mazzucato2019expectation}, as opposed to the contrast response function. 
We have classified mechanisms of gain modulation which act by changing the mean or spatial variance {\it across neurons} of the cell-type specific afferent currents to the local cortical circuit, where we modeled afferent currents as constant biases; or by changing the recurrent couplings. The rationale for our choice was to investigate the effects on internally generated variability in a network whose dynamics were entirely deterministic. Alternatively, one could model external currents as time-dependent inputs with fast noise, such as Poisson processes or colored noise. In that case, changes in background noise due to barrages of synaptic inputs are capable of inducing gain modulation as well \cite{chance2002gain,haider2009rapid}. Previous work compared these different kinds of perturbations (Poisson noise or afferent spatial variance) in the case of the perturbation $\delta$var(E) \cite{mazzucato2019expectation}, showing they may lead to similar outcomes.

\subsection{Physiological mechanisms of gain modulation}

Several different physiological pathways can modulate the gain of the intrinsic neuronal transfer function, including neuromodulation, top-down and cortico-cortical interactions. Gain modulation can also be induced artificially by means of optogenetic or pharmacological manipulations. The perturbations investigated in our model may be related to different pathways and implicated in various types of cognitive function.

\subsubsection*{Neuromodulation}

Neuromodulatory pathways strongly affect sensory processing in cortical circuits by changing cell-type specific afferent currents to the circuit, in some cases controlling their dynamical regime \cite{mcginley2015waking}. Our theory may be applicable to explain the effects of cholinergic and serotonergic activation on sensory cortex.

Cholinergic pathways, modulating ionic currents in pyramidal cells \cite{mccormick1992neurotransmitter}, can control cortical states and mediate the effects of arousal and locomotion. Artificial stimulation of cholinergic pathways was shown to improve sensory coding in visual  \cite{goard2009basal,pinto2013fast} and barrel cortex \cite{eggermann2014cholinergic}. Cholinergic stimulation alone in the absence of sensory stimuli was shown to induce mixed responses with different neural populations increasing or decreasing their spiking activity  \cite{goard2009basal}. Our theory shows that these combined experimental observations (coding improvement and mixed firing rate changes) are consistent with a mechanism whereby cholinergic activation induces an increase in $\delta$var(E) afferents to sensory cortex, inducing an acceleration of sensory processing (Fig. \ref{figtwo}e and S1a).

Activation of serotonergic pathways by stimulation of dorsal raphe serotonergic neurons or local iontophoresis was shown to transiently degrade stimulus coding in sensory cortex, decreasing responses to mechanosensory stimuli \cite{dugue2014optogenetic} and increasing the latency of the first spike evoked by auditory stimuli \cite{hurley2002serotonin}. Serotonergic stimulation was shown to decrease firing rates in the olfactory cortex \cite{lottem2016optogenetic}, inferior colliculus \cite{hurley2002serotonin}, and primary visual cortex \cite{michaiel2019hallucinogenic, seillier2017serotonin}. Our theory shows that these experimental observations (coding degradation and decreased firing rates) are consistent with two alternative mechanisms (Fig. \ref{figtwo}d and S1a): either an increase in the afferent currents to I populations (i.e.,  $\delta$mean(I)$>0$) implementing the paradoxical inhibition effect \cite{tsodyks1997paradoxical}; or  a decrease in the afferents to E populations (i.e.,  $\delta$mean(E)$<0$). Future experiments could test between these two alternatives.

\subsubsection*{Top-down projections} 

A prominent feature of sensory cortex is the integration of feedforward and cortico-cortical feedback pathways at each stage of sensory processing \cite{felleman1991distributed}. In particular, top-down projections from higher cortical areas to sensory cortex are known to modulate the speed and accuracy of sensory processing \cite{mazzucato2019expectation}. Our theory may explain the effects of activating several cortico-cortical pathways.

Activation of feedback axons from motor cortex (M1) to somatosensory cortex (S1) was shown to increase activity in S1 during whisking \cite{petreanu2012activity} and led to faster and more accurate responses to whisker stimulation \cite{zagha2013motor}. Suppression of the same pathway induced slower S1 responses to whisking in awake mice. Our theory shows that the effect of these cortico-cortical perturbations is consistent with an increase in the mean afferent currents to E populations in S1 (i.e., the $\delta$mean(E) perturbation in Fig. \ref{figtwo}d), leading to higher firing rates and faster processing speed. 

Expectation and arousal are known to strongly modulate neural activity in sensory cortices \cite{salkoff2020movement}. Expected stimuli are processed faster and more accurately than unexpected stimuli both in auditory \cite{jaramillo2010auditory} and gustatory cortex \cite{samuelsen2012effects}. Experimental evidence shows that the anticipation of sensory processing induced by expectation is mediated by top-down projections from the amygdala to the gustatory cortex \cite{samuelsen2012effects}, whose activation elicits mixed excited and inhibited responses in both pyramidal and inhibitory cells in the gustatory cortex \cite{samuelsen2012effects,vincis2016associative}. Our model shows that, while an acceleration of stimulus processing speed may in principle be mediated by different state-changing perturbations, only the $\delta$var(E) perturbation is consistent with the empirically observed mixed responses. Indeed, our theory suggests that these top-down projections may operate by inducing an increase in the spatial variance of the  afferent currents to the E population \cite[$\delta$var(E) in Fig. \ref{figtwo}d, extending previous results in][to networks including inhibitory clusters]{mazzucato2019expectation} .

In attentional tasks, distractors slow down reaction times \cite{grueninger1969effects,treisman1980feature}, a behavioral effect that may be mediated by changes in the speed and accuracy of sensory processing in cortical circuits \cite{desimone1995neural}. The presence of distracting stimuli within a neurons receptive field suppresses its responses to the preferred stimulus \cite{knierim1992neuronal}. The underlying mechanism may recruit lateral inhibition onto the local cortical circuit \cite{reynolds2009normalization,gilbert2013top}. Our theory shows that this mechanism is consistent with a modulation of the afferents to local I populations, mediated by either an increase in $\delta$mean(I) or $\delta$var(I) in Fig. \ref{figtwo}d. It would be interesting to discriminate between these two perturbations with future experiments.

\subsubsection*{Optogenetic and pharmacological manipulations}

Our theory may shed light on the effects of manipulation experiments. Optogenetic activation (inactivation) of specific E or I cells \cite{arenkiel2007vivo,li2019spatiotemporal} has been modeled as an increase (decrease) of the afferent currents to those cells \cite{ebsch2018imbalanced,mahrach2020mechanisms,sanzeni2019inhibition}. However, protein expression may not be complete across all cells of the targeted population, and even in the case of complete expression across the targeted population, different cells may be more or less sensitive to laser stimulation. Thus the effect of optogenetic stimulation on the targeted population may then be more accurately modeled by a concurrent change in both mean and variance of the targeted cell-type specific afferents (e.g., $\delta$mean(E) and $\delta$var(E) for E populations; $\delta$mean(I) and $\delta$var(I) for I populations). Recent studies showed that, while a homogeneous stimulation of all I cell types simultaneously can be captured by a model of E-I recurrently coupled neurons (as in our model), partial activation of specific inhibitory cell-types may induced more complex responses \cite{mahrach2020mechanisms,sanzeni2019inhibition,li2019spatiotemporal,otchy2015acute,zur2008direct,phillips2016asymmetric}. We plan to revisit this issue in the future.  

Our theory may also be applicable to the effects of pharmacological manipulations of different synaptic receptors. In particular, the effects of combined local injection of AMPA/kainate and NMDA receptor antagonists (agonists) may be recapitulated by a decrease (increase) in $\delta$AMPA, which correspondingly perturb the value of $J_{IE},J_{EE}$ couplings (Fig. \ref{figtwo}d). Similarly, the effects of local injection of GABA receptor antagonists (agonists) may be recapitulated by a decrease (increase) in $\delta$GABA, which correspondingly perturb the value of $J_{EI},J_{II}$ couplings.


\subsection{Locomotion and gain modulation}

Locomotion has been shown to modulate visually evoked activity \cite{niell2010modulation} and is sufficient in driving activity in mouse V1 \cite{leinweber2017sensorimotor,saleem2013integration}. Our results were consistent with previous studies in showing that locomotion affects the activity of neurons in the visual cortical hierarchy during both ongoing and stimulus-evoked activity. We found that locomotion in the absence of sensory stimuli induces an average increase in firing rates. At the single-cell level we reported a complex mix of excited and inhibited responses in both E and I cells, also consistent with previous results \cite{fu2014cortical,dipoppa2018vision}. Crucially, we uncovered that locomotion decreased the single-cell gain \com{ during ongoing activity across the board in the visual cortical hierarchy (Fig. \ref{figseven}d). Our theory predicted that the observed decrease in gain would lead to an acceleration of visual processing during locomotion in cortex. This prediction was confirmed in the data (Fig. \ref{figseven}i).} The acceleration of processing speed \com{observed in cortex} did not depend on the locomotion-induced changes in firing rates and was still present even after matching the firing rate distributions between running and rest conditions (Fig. \ref{figsfive}). \com{Our model of the perturbation effects induced by locomotion} (increased firing rates with mixed excited and inhibited responses, and faster visual processing) suggests that the effect of locomotion may be mediated by a increase in the spatial variance of the afferent current to the E populations ($\delta$var(E) perturbation) \cite{ayaz2013locomotion,niell2010modulation,fu2014cortical}. Concretely, gain modulation may be implemented via the combined effect of activating neuromodulatory pathways such as cholinergic \cite{fu2014cortical} and noradrenergic \cite{polack2013cellular} inputs.

\section{Acknowledgements}

This work was supported by a National Institute of Deafness and Other Communication Disorders Grant K25-DC013557 and a National Institute of Neurological Disorders and Stroke Grant NS118461 (BRAIN Initiative). We would like to thank G. La Camera, Y. Ahmadian, and M. Wehr for useful discussions and suggestions.

\bibliographystyle{unsrt}
\bibliography{bib}

\newpage
\newcommand{\beginsupplement}{%
        \setcounter{table}{0}
        \renewcommand{\thetable}{S\arabic{table}}%
        \setcounter{figure}{0}
        \renewcommand{\thefigure}{S\arabic{figure}}%
     }
\beginsupplement

\renewcommand{\thefigure}{2-1}
\begin{figure}[ht]
\hspace{-1cm}
\includegraphics[width=0.8\textwidth]{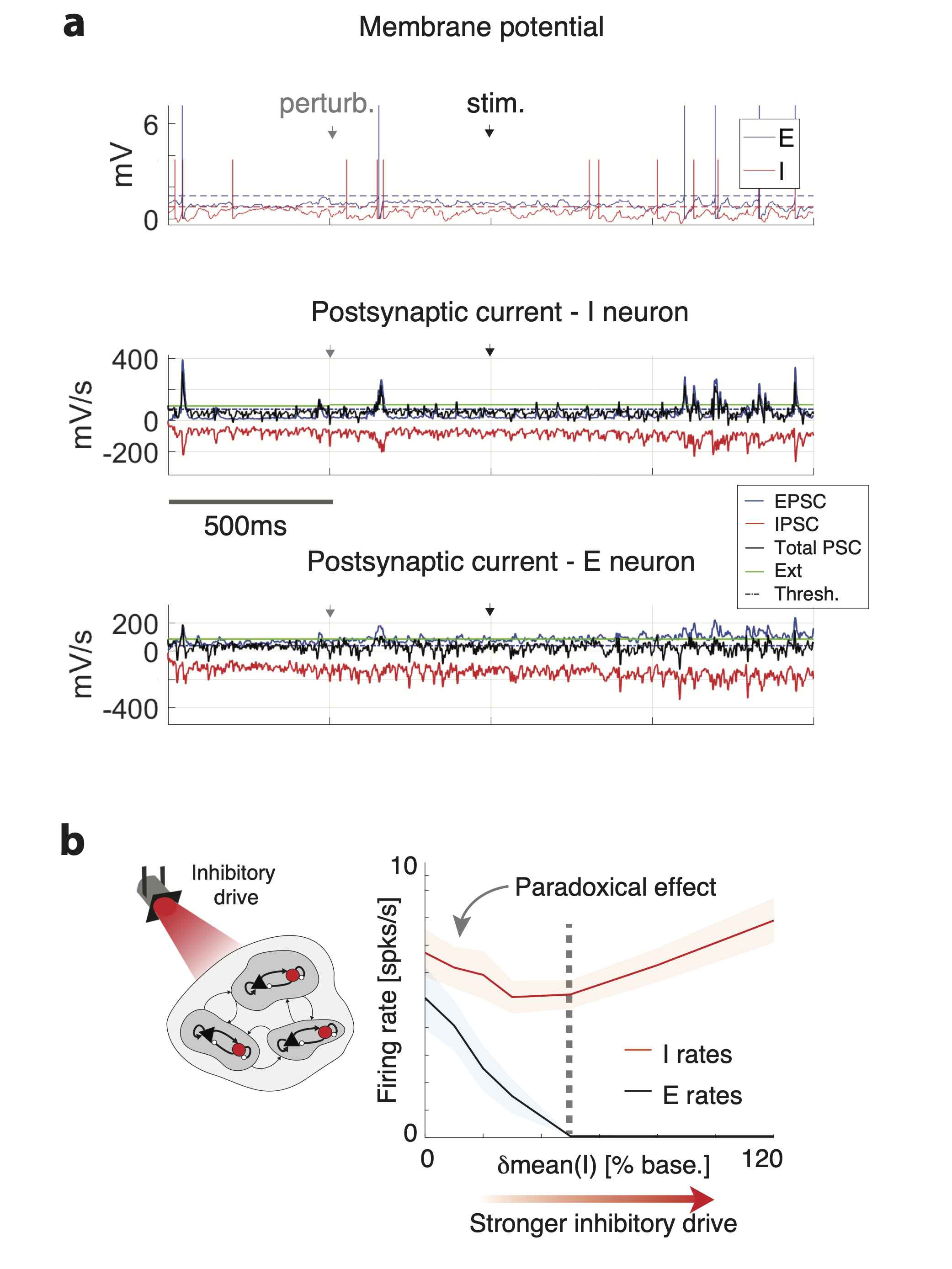}
\vspace{-0cm}
\caption{\com{Inhibition stabilized balanced regime. {\bf a}) Top: Membrane potential from representative E (blue) and I (red) neurons (vertical bars: action potentials; horizontal dashed lines: threshold for spike emission; grey and black arrows represent state-changing $\delta$var(E) perturbation onset and sensory stimulus onset, respectively; same network as in \ref{figfour}b). Incoming post-synaptic current (PSC) to an I (center) and an E (bottom) neuron: EPSC (blue trace), IPSC (red trace), external current (green line), and total current (black trace) are in a tight balanced regime. {\bf b}) When increasing the inhibitory drive (afferent current to the I population, same as $\delta$mean(I) perturbation), both E and I firing rates decrease (black and red curve in right panel, mean$\pm$s.e.m. across 10 simulated networks), highlighting the paradoxical effect, signature of the inhibition stabilized regime \cite{tsodyks1997paradoxical}. Beyond $\delta$mean(I)=$50\%$ the E population shuts down and the I population rebounds (dashed vertical line). }
}
\label{figsone}
\end{figure}


\renewcommand{\thefigure}{3-1}
\begin{figure}[ht]
\vspace{-1cm}
\includegraphics[width=1.\textwidth]{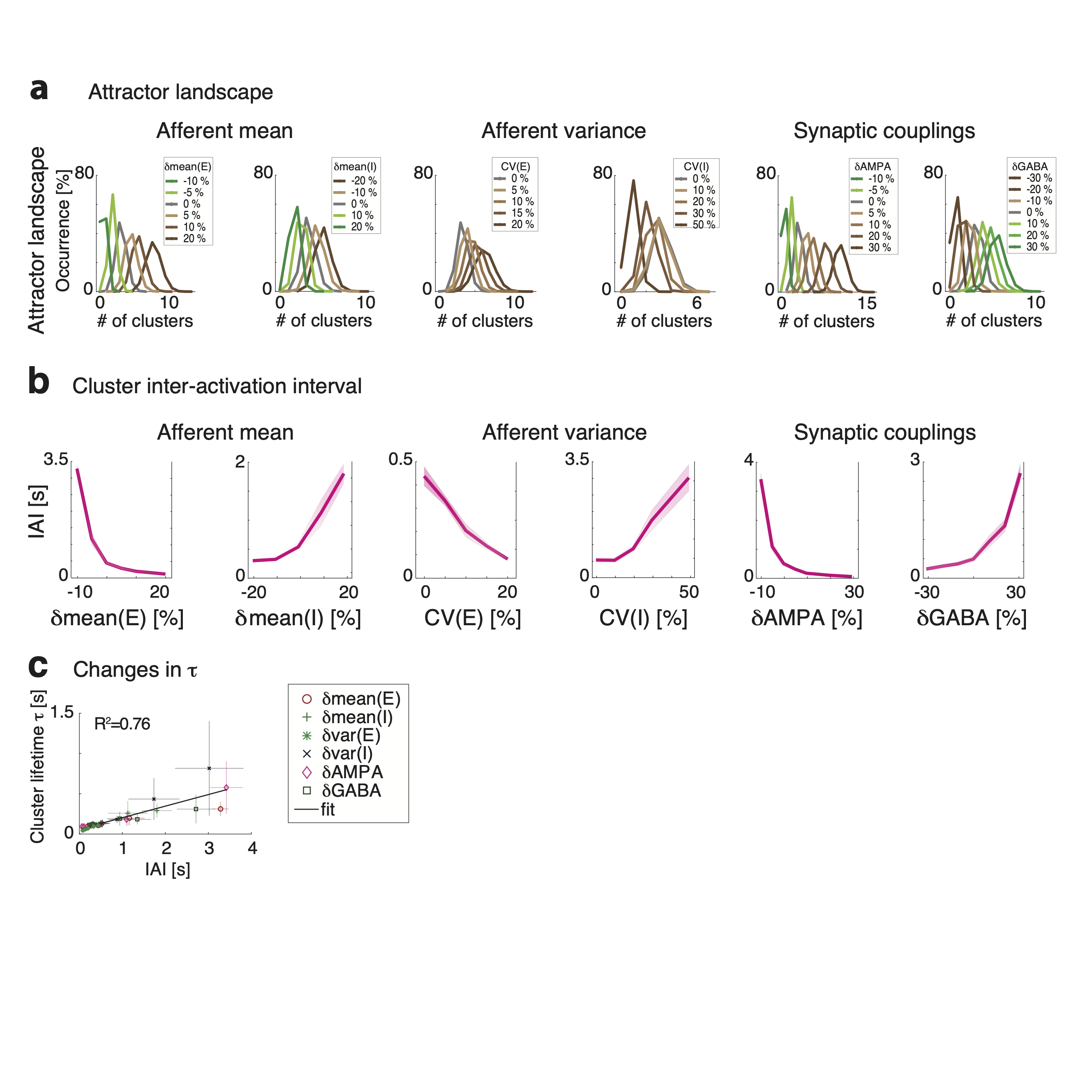}
\vspace{-2cm}
\caption{{\bf a}) Perturbations modulate the attractor landscape (color-coded curves: frequency of occurrence of network attractors with specified number of activated clusters, for each value of the perturbation, mean across 5 sessions; notations as in \ref{figthree}a). Perturbations-induced modulations of timescales. Perturbation-induced changes in the cluster inter-activation interval (IAI, {\bf b}) closely track the changes in cluster activation lifetime $\tau$ ({\bf c}).}
\label{figstwo}
\end{figure}


\renewcommand{\thefigure}{3-2}
\begin{figure}[ht]
\vspace{-1cm}
\includegraphics[width=1.\textwidth]{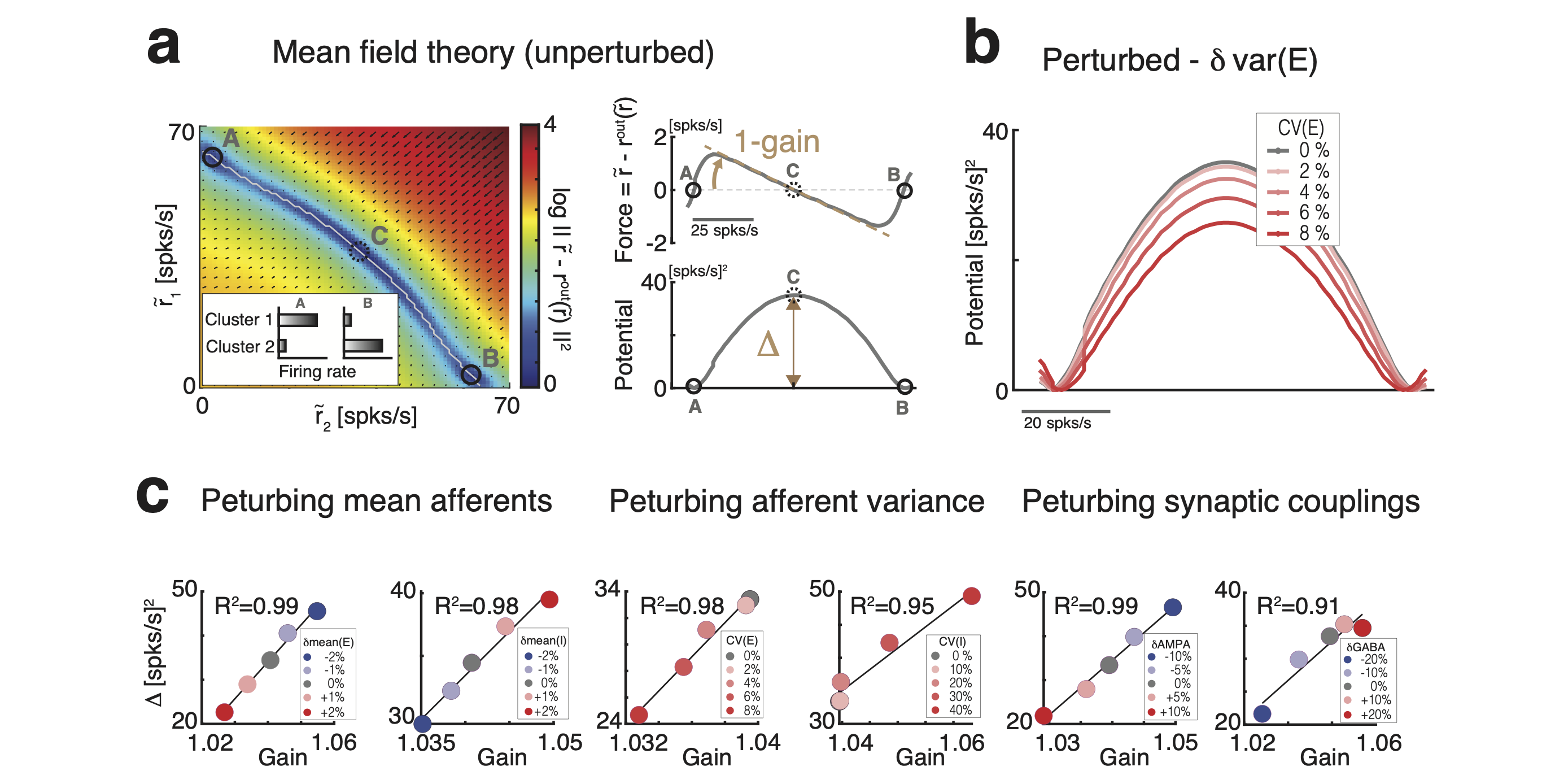}
\vspace{-0cm}
\caption{Effective mean field theory for barrier height and gain. {\bf a}) Left: Reduced two-cluster theory showing the force vector (black arrows, color-coded map represents the log of the force vector norm) acting on a configuration where the two clusters have firing rates $\tilde r_1,\tilde r_2$. The force vanishes at the stable fixed points $A$ and $B$, corresponding to attractors where either cluster is active and the other inactive (inset), and at saddle point $C$ between them. Top right: From the projection of the force vector on the trajectory between the attractors (white curve in left panel) one obtains an effective transfer function $r^{out}(\tilde r)$ whose slope yields the population intrinsic gain. Bottom right: The energy barrier separating the two attractors A and B is defined as the line integral of the projected force along the trajectory. {\bf b}): The perturbation $\delta$var(E) lowers the energy barrier between the two attractors (darker color-shades represent increasing values CV(E) of the perturbation). {\bf c}) Mean field theory predicts a direct relationship between the height of the barrier $\Delta$ separating the attractors and the gain for all perturbations.
}
\label{figsthree}
\end{figure}


\renewcommand{\thefigure}{6-1}
\begin{figure}[ht]
\vspace{-1cm}
\hspace{-0.5cm}
\includegraphics[width=1.\textwidth]{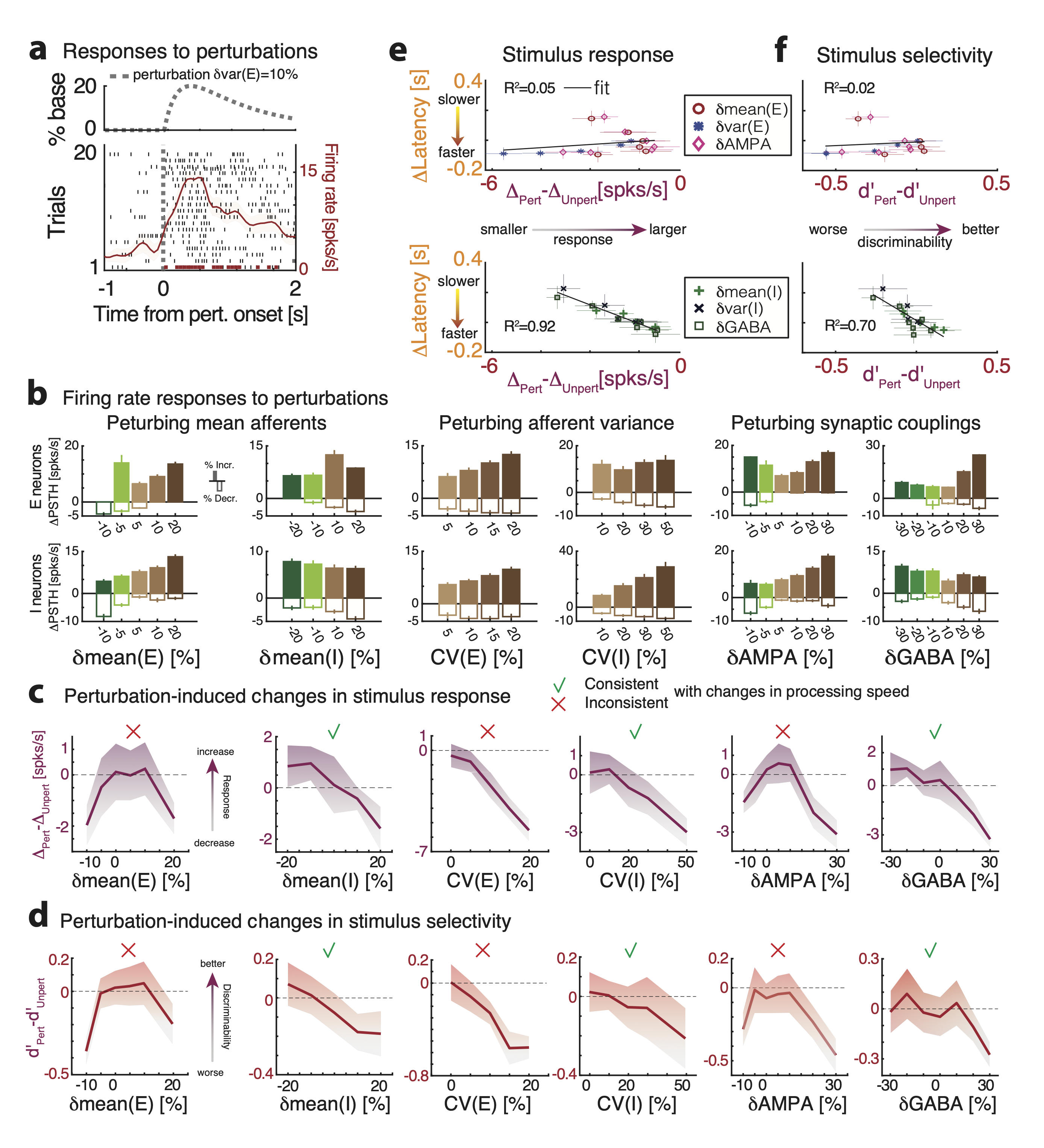}
\vspace{-0cm}
\caption{Perturbation-induced changes in single-cell response. {\bf a}) Representative single cell response to the perturbation $\delta$var(E)=10$\%$ in the absence of stimuli (top: dashed line, time course of perturbation, occurring at $t=0$; bottom: dashed line, perturbation onset; red curve, response PSTH, mean$\pm$S.D. across 20 trials; horizontal red bar: significant response, t-test, $p<0.05$ with multiple bin correction. {\bf b}) Changes in peak-firing rate compared to baseline ($\Delta$PSTH=peak-baseline; positive for firing rate increase, negative decrease) for E and I neurons (only neurons responsive to the state-changing perturbations, fractions reported in Fig. \ref{figthree}b), in response to a perturbation with time course as in panel a. {\bf c}: Single-cell changes in stimulus selectivity due to the perturbations: d'(perturbed trials)-d'(unperturbed trials).
{\bf e}) Single-cell changes in firing rate response to stimuli due to the perturbations ($\Delta=$peak response-baseline in each perturbed or unperturbed condition) are overall uncorrelated to changes in stimulus-decoding latency (mean$\pm$s.e.m. across 5 networks; same as panel b). {\bf f}) Single-cell changes in stimulus selectivity due to the perturbations (d') are overall uncorrelated to changes in stimulus-decoding latencies (same notation as in panel c).}
\label{figsfour}
\end{figure}


\renewcommand{\thefigure}{7-1}
\begin{figure}[ht]
\vspace{-1cm}
\includegraphics[width=1\textwidth]{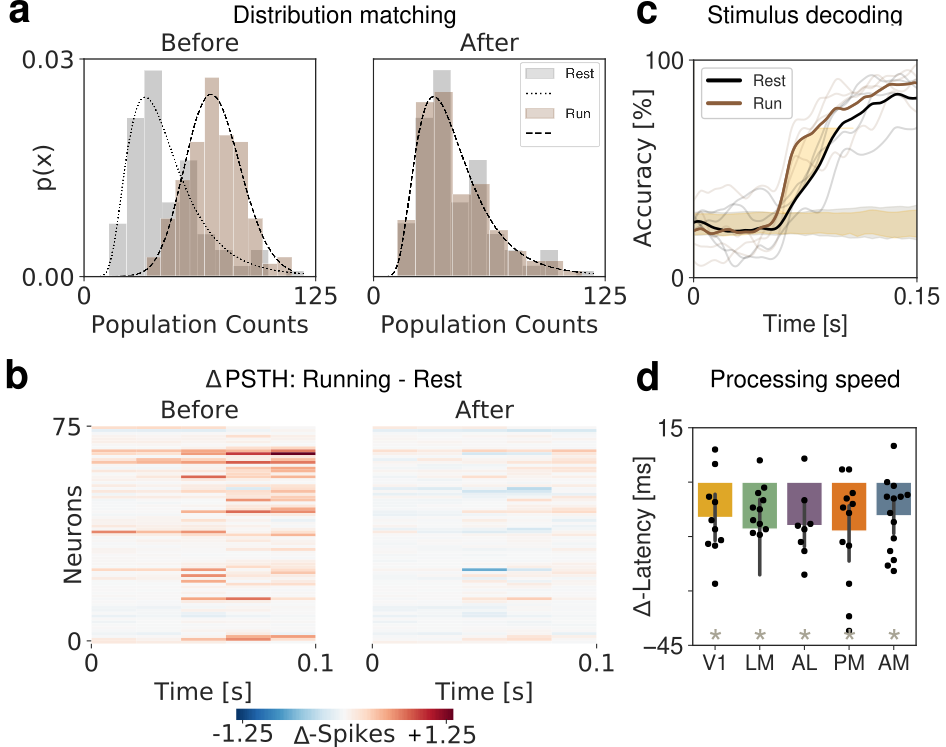}
\vspace{-1cm}
\caption{Anticipation of stimulus decoding persists even after matching the distribution of firing rates across behavioral conditions. {\bf a}) Firing rate distributions for both rest and running before (left) and after (right) randomly removing spikes from the running condition. Black lines show log-normal fits of distributions. {\bf b}) {$\Delta$} PSTH between behavioral conditions before and after distribution matching shows effects of match across each neuron's firing rate. {\bf c}) Mean stimulus-decoding accuracy across orientations per behavioral condition using neurons from V1 as predictors shows the anticipation of the stimulus in the running condition after distribution matching (same sessions as in Fig. \ref{figsix}e). {\bf d}) Summary of changes in processing speed due to locomotion by area after distribution matching. (t-test, p< 0.01)}
\label{figsfive}
\end{figure}


\renewcommand{\thefigure}{7-2}
\begin{figure}[ht]
\vspace{-1cm}
\hspace{1cm}
\includegraphics[width=0.8\textwidth]{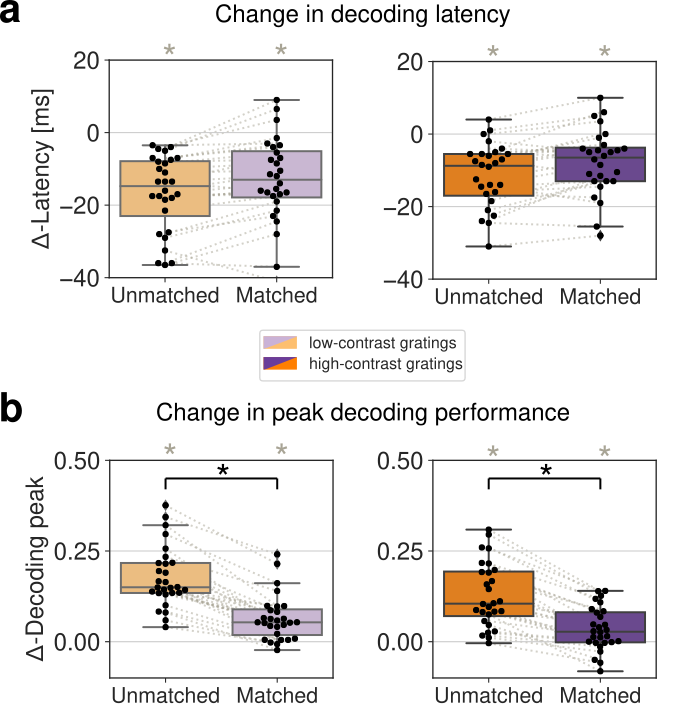}
\vspace{0cm}
\caption{Matching the distribution of firing rates between behavioral conditions reduces the change in peak decoding, but preserves the change in decoding latency between behavioral conditions. {\bf a}) {$\Delta$}-Latency over all areas, separated by the grating contrast shows that even after matching the distribution of firing rates between conditions (purple), the increase in sensory processing during running was still significant (rank-sum test, gray $*=p <0.005$) {\bf b}) The difference in peak decoding between behavioral conditions is reduced for low and high contrast drifting grating trials after matching the distributions (rank-sum test, gray $*=p <0.005$). The change in {$\Delta$}-Decoding peak between non-matched and matched  datasets was significant (rank-sum test, black $*=p <0.005$) for both contrasts. }
\label{figssix}
\end{figure}


\renewcommand{\thefigure}{7-3}
\begin{figure}[ht]
\vspace{-1cm}
\includegraphics[width=0.8\textwidth]{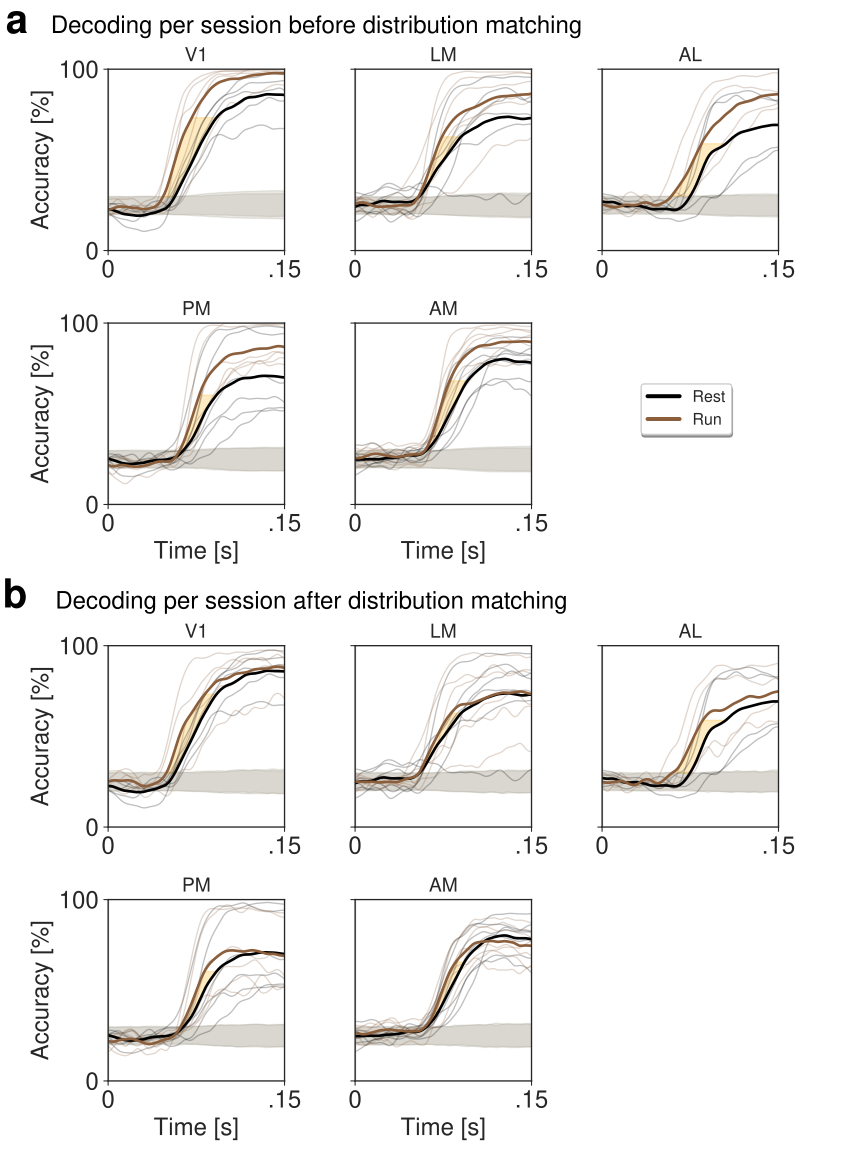}
\vspace{0cm}
\caption{Mean stimulus-decoding accuracy of high-contrast drifting gratings across sessions per behavioral condition and area before ({\bf a}) and after ({\bf b}) matching the distribution of firing rates shows the decrease in {$\Delta$}-Decoding peaks and preservation of {$\Delta$}-Latency. Notations as in Fig. \ref{figsix}e.}
\label{figsseven}
\end{figure}


\renewcommand{\thefigure}{7-4}
\begin{figure}[ht]
\vspace{-1cm}
\includegraphics[width=0.8\textwidth]{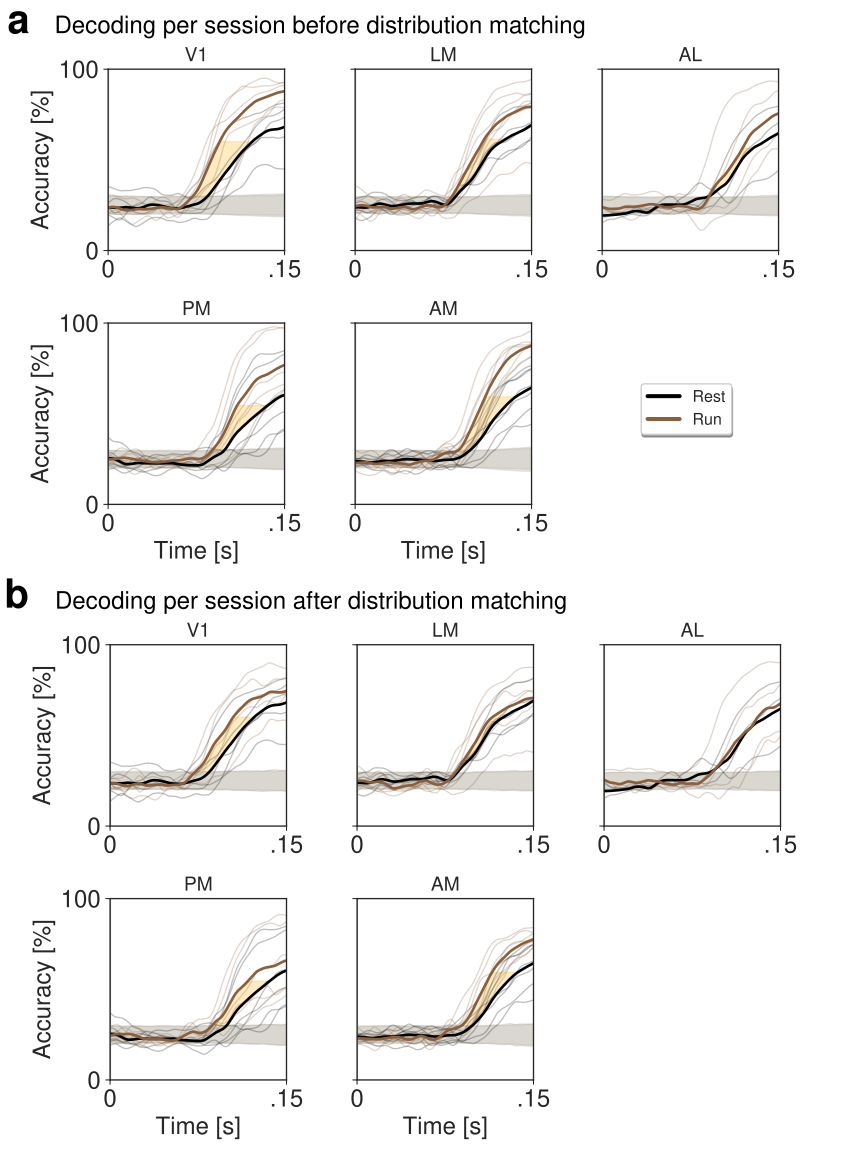}
\vspace{0cm}
\caption{Mean stimulus-decoding accuracy of low-contrast drifting gratings across sessions per behavioral condition and area before ({\bf a}) and after ({\bf b}) matching the distribution of firing rates shows the decrease in {$\Delta$}-Decoding peaks and preservation of {$\Delta$}-Latency. Notations as in Fig. \ref{figsix}e.}
\label{figseight}
\end{figure}


\renewcommand{\thefigure}{7-5}
\begin{figure}[!p]
\vspace{-1cm}
\includegraphics[width=1\textwidth]{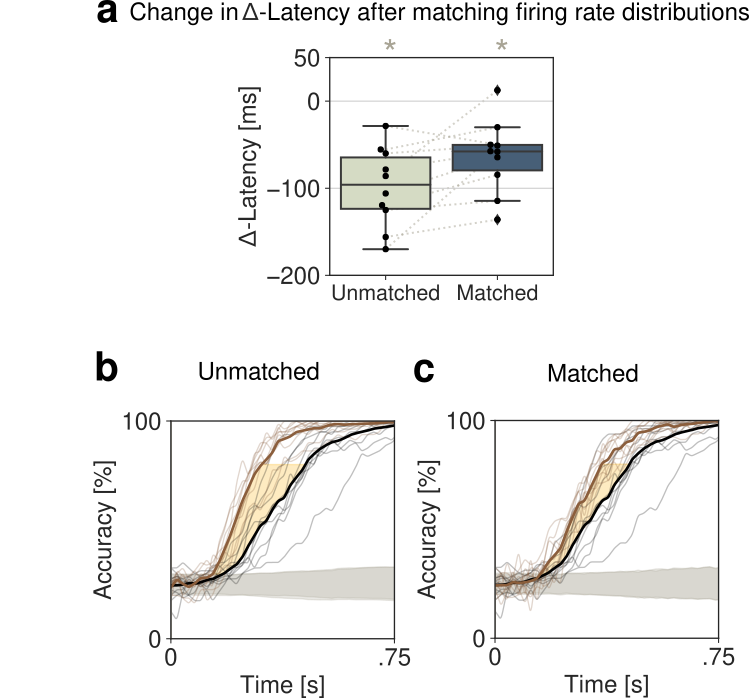}
\vspace{0cm}
\caption{In the model, matching the distribution of firing rates between perturbed ($\delta$var(E) with CV(E)=20\%) and unperturbed conditions preserved the perturbation-induced acceleration in stimulus processing speed (same data as in Fig. \ref{figtwo}b). {\bf a}) {$\Delta$}-Latency over 10 simulated networks shows that even after matching the distribution of firing rates between conditions (purple), the increase in sensory processing speed during the perturbed condition was still significant (rank-sum test, $*=p<0.005$). There was no significant change in {$\Delta$}-Latency between unmatched and matched datasets (rank-sum test,$p >0.05$). Time course of stimulus-decoding accuracy over all 10 simulated networks  before ({\bf b}) and after ({\bf c}) matching the distribution of firing rates.}
\label{figsnine}
\end{figure}



\end{document}